\documentclass[twocolumn]{aastex7}
\usepackage{wrapfig}
\usepackage{amsmath}
\usepackage{amssymb}
\usepackage{rotating}
\usepackage{epstopdf}
\usepackage{makecell}
\usepackage{comment}

\usepackage[flushleft]{threeparttable}

\PassOptionsToPackage{hyphens}{url}\usepackage{hyperref}

\newcommand{\snana}{\textsc{snana}}

\newcommand{\nSNeAbout}{\textcolor{black}30\,000}

\newcommand{\nIasnrFive}{\textcolor{black}{13\,000}}
\newcommand{\nIasnrThree}{\textcolor{black}{18\,000}}
\newcommand{\nCCThree}{\textcolor{black}{9\,000}}
\newcommand{\nSLSNThree}{\textcolor{black}{3\,000}}
\newcommand{\nfaintandFastThree}{\textcolor{black}{500}}

\newcommand{\nGotHosts}{\textcolor{black}{400\,000}}
\newcommand{\nHostsSNRthree}{\textcolor{black}{200\,000}}
\newcommand{\nHostsIASNRthree}{\textcolor{black}{135\,000}}

\newcommand{\nAllCosmo}{\textcolor{black}{143\,000}}
\newcommand{\nAllCosmoAbout}{\textcolor{black}{140\,000}}
\newcommand{\nCosmoLiveIa}{\textcolor{black}{12\,600}}
\newcommand{\nCosmoHosts}{\textcolor{black}{131\,000}}

\newcommand{\snrF}{SNR$_{15\text{\AA}}$\,}
\newcommand{\snrO}{SNR$_{1\text{\AA}}$\,}
\defcitealias{2019MNRAS.489.5802V}{V19}
\defcitealias{2021MNRAS.505.2819V}{V21}

\begin{document}

\title{TiDES: The 4MOST Time Domain Extragalactic Survey}

\author[0000-0001-9553-4723]{C. Frohmaier}
\affiliation{School of Physics and Astronomy, University of Southampton, Southampton, SO17 1BJ, UK}
\email[show]{c.frohmaier@soton.ac.uk}
\correspondingauthor{C. Frohmaier}

\author[0000-0001-8788-1688]{M. Vincenzi}
\affiliation{Department of Physics, University of Oxford, Keble Road, Oxford, OX1 3RH, UK}
\email{maria.vincenzi@physics.ox.ac.uk}

\author[0000-0001-9053-4820]{M. Sullivan}
\affiliation{School of Physics and Astronomy, University of Southampton, Southampton, SO17 1BJ, UK}
\email{m.sullivan@soton.ac.uk}

\author[0000-0002-6353-1111]{S.F. H\"onig}
\affiliation{School of Physics and Astronomy, University of Southampton, Southampton, SO17 1BJ, UK}
\email{s.hoenig@soton.ac.uk}

\author[0000-0002-3321-1432]{M. Smith}
\affiliation{Department of Physics, Lancaster University, Lancaster, LA1 4YB, UK}
\email{mat.smith@soton.ac.uk}

\author[0000-0001-8271-1364]{H. Addison}
\affiliation{School of Mathematics and Physics, University of Surrey, Guildford, GU2 7XH}
\email{ha00871@surrey.ac.uk}

\author[0000-0001-5564-3140]{T. Collett}
\affiliation{Institute of Cosmology and Gravitation, University of Portsmouth, Portsmouth PO1 3FX, UK}
\email{thomas.collett@port.ac.uk}

\author[0000-0001-9494-179X]{G. Dimitriadis}
\affiliation{Department of Physics, Lancaster University, Lancaster, LA1 4YB, UK}
\email{g.dimitriadis@lancaster.ac.uk}

\author[0000-0001-7782-7071]{R. S. Ellis}
\affiliation{Department of Physics \& Astronomy, University College London, London WC1E 6BT, UK}
\email{richard.ellis@ucl.ac.uk}

\author[0000-0003-3105-2615]{P. Gandhi}
\affiliation{School of Physics and Astronomy, University of Southampton, Southampton, SO17 1BJ, UK}
\email{poshak.gandhi@soton.ac.uk}

\author[0000-0002-4391-6137]{O. Graur}
\affiliation{Institute of Cosmology and Gravitation, University of Portsmouth, Portsmouth PO1 3FX, UK}
\affiliation{Department of Astrophysics, American Museum of Natural History, Central Park West and 79th Street, New York, NY 10024-5192, USA}
\email{or.graur@port.ac.uk}

\author[0000-0002-2960-978X]{I. Hook}
\affiliation{Department of Physics, Lancaster University, Lancaster, LA1 4YB, UK}
\email{i.hook@lancaster.ac.uk}

\author[0000-0003-0313-0487]{L. Kelsey}
\affiliation{Institute of Astronomy and Kavli Institute for Cosmology, University of Cambridge, Madingley Road, Cambridge CB3 0HA, UK}
\email{lisa.kelsey@ast.cam.ac.uk}

\author[0000-0002-1031-0796]{Y.-L. Kim}
\affiliation{Department of Physics, Lancaster University, Lancaster, LA1 4YB, UK}
\affiliation{Department of Astronomy \& Center for Galaxy Evolution Research, Yonsei University, Seoul 03722, Republic of Korea}
\email{olo8308@gmail.com}

\author[0000-0003-1731-0497]{C. Lidman}
\affiliation{The Research School of Astronomy and Astrophysics, Australian National University, ACT 2601, Australia}
\affiliation{Centre for Gravitational Astrophysics, College of Science, The Australian National University, ACT 2601, Australia}
\email{christopher.lidman@anu.edu.au}

\author[0000-0002-9770-3508]{K. Maguire}
\affiliation{School of Physics, Trinity College Dublin, The University of Dublin, Dublin 2, Ireland}
\email{kate.maguire@tcd.ie}

\author[0000-0002-7466-4868]{L. Makrygianni}
\affiliation{Department of Physics, Lancaster University, Lancaster, LA1 4YB, UK}
\email{l.makrygianni@lancaster.ac.uk}

\author[0009-0006-4963-3206]{B. Martin}
\affiliation{The Research School of Astronomy and Astrophysics, Australian National University, ACT 2601, Australia}
\email{bailey.martin@anu.edu.au}

\author[0000-0001-8211-8608]{A. M\"oller}
\affiliation{Centre for Astrophysics \& Supercomputing, Swinburne University of Technology, Victoria 3122, Australia}
\email{amoller@swin.edu.au}

\author[0000-0003-0939-6518]{R. C. Nichol}
\affiliation{School of Mathematics and Physics, University of Surrey, Guildford, GU2 7XH}
\email{bob.nichol@surrey.ac.uk}

\author[0000-0002-2555-3192]{M. Nicholl}
\affiliation{Astrophysics Research Centre, School of Mathematics and Physics, Queens University Belfast, Belfast BT7 1NN, UK}
\email{matt.nicholl@qub.ac.uk}

\author[0000-0002-1214-770X]{P. Schady}
\affiliation{Department of Physics, University of Bath, Claverton Down, Bath, BA2 7AY, UK}
\email{ps2018@bath.ac.uk}

\author[0000-0001-5882-3323]{B. D. Simmons}
\affiliation{Department of Physics, Lancaster University, Lancaster, LA1 4YB, UK}
\email{b.simmons@lancaster.ac.uk}

\author[0000-0002-8229-1731]{S. J. Smartt}
\affiliation{Department of Physics, University of Oxford, Keble Road, Oxford, OX1 3RH, UK}
\affiliation{Astrophysics Research Centre, School of Mathematics and Physics, Queens University Belfast, Belfast BT7 1NN, UK}
\email{stephen.smartt@physics.ox.ac.uk}

\author[0000-0002-5249-7018]{E. Tempel}
\affiliation{Tartu Observatory, University of Tartu, Observatooriumi 1, 61602 Tõravere, Estonia}
\email{elmo.tempel@ut.ee}

\author[0000-0002-3073-1512]{P. Wiseman}
\affiliation{School of Physics and Astronomy, University of Southampton, Southampton, SO17 1BJ, UK}
\email{p.s.wiseman@soton.ac.uk}

\collaboration{all}{and the LSST Dark Energy Science Collaboration}


\begin{abstract}

The Time Domain Extragalactic Survey (TiDES) conducted on the 4-metre Multi-Object Spectroscopic Telescope (4MOST) will perform spectroscopic follow-up of extragalactic transients discovered in the era of the NSF-DOE Vera C. Rubin Observatory. TiDES will conduct a 5-year survey, covering ${>}14\,000\,\mathrm{square\, degrees}$, and use around 250 000 fibre hours to address three main science goals: (i) spectroscopic observations of ${>}$\nSNeAbout\ live transients, (ii) comprehensive follow-up of ${>}$\nHostsSNRthree\ host galaxies to obtain redshift measurements, and (iii) repeat spectroscopic observations of Active Galactic Nuclei to enable reverberation mapping studies. The live spectra from TiDES will be used to reveal the diversity and astrophysics of both normal and exotic supernovae across the luminosity-timescale plane. The extensive host-galaxy redshift campaign will allow exploitation of the larger sample of supernovae and improve photometric classification, providing the largest-ever sample of type Ia supernovae, capable of a sub-2 per cent measurement of the equation-of-state of dark energy. Finally, the TiDES reverberation mapping experiment of 700--1\,000 AGN will complement the SN Ia sample and extend the Hubble diagram to $z\sim2.5$.

\end{abstract}




\section{Introduction}

The NSF-DOE Vera C. Rubin Observatory's Legacy Survey of Space and Time \citep[LSST;][]{2019ApJ...873..111I} heralds a new era of transient discovery: LSST will produce millions of transient alerts and photometric data for hundreds-of-thousands of supernovae (SNe) and other variable sources. While this discovery rate is exciting, the ability to obtain spectroscopic follow-up of these transients must similarly scale to enable much of the scientific exploitation. Follow-up spectra are critical for extracting the full astrophysical potential of the objects discovered, for example their classifications, rates, chemistry, redshifts, luminosities -- and ultimately their physical nature.
The capacity to assemble these spectroscopic samples presents a new challenge for the transient community, and to meet this, a next generation facility is needed.

In this paper we present the Time Domain Extragalactic Survey (TiDES), which is designed to meet the spectroscopic requirements of the LSST-era for optical transients, their host galaxies, and active galactic nuclei (AGN). TiDES will use the 4-metre Multi-Object Spectroscopic Telescope \citep[4MOST;][]{2019Msngr.175....3D}, a fibre-fed spectroscopic survey facility on the European Southern Observatory (ESO) Visible and Infrared Survey Telescope for Astronomy (VISTA). This facility packs $2,436$ science fibres distributed over the 4.2\,deg$^2$ field-of-view, with each spine configurable by the AESOP fibre positioning system onto a target of interest. Throughout this work we refer to the concept of a fibre-hour. A fibre-hour is one hour of observing time, including overheads, for one fibre. In total, 4MOST provides up to 2,436 fibre-hours per one hour of observation. TiDES will exploit the fact that wherever 4MOST points over ${\sim}18\,000$\,deg$^2$ of the extragalactic sky, there will be known time-variable sources recently discovered by Rubin or other surveys. These will include recently discovered transients, and older, faded events for which host galaxy spectroscopy will be obtained for measuring redshifts. The strategy takes inspiration from the successful \lq OzDES\rq\ survey \citep[][]{2015MNRAS.452.3047Y,2020MNRAS.496...19L}, conducted using the 2dF fibre positioner and the AAOmega spectrograph on the Anglo-Australian Telescope, which secured live SN spectra and ${\sim}$4,500 redshifts \citep{2019MNRAS.489.5802V} for SN host galaxies over 27\,deg$^2$ of imaging from the Dark Energy Survey \citep[DES;][]{2016MNRAS.460.1270D}.

TiDES will use around 250\,000 fibre-hours to tackle three key science goals. The first is the use of type Ia supernovae (SNe Ia) as a cosmological probe. SNe Ia are an exceptionally mature and well-understood cosmological probe via their use as standardisable candles \citep[][]{1998AJ....116.1009R, 1999ApJ...517..565P,2014A&A...568A..22B,2024ApJ...973L..14D}. They remain a uniquely powerful distance-indicator in the $z<0.5$ Universe, and directly constrain the properties of dark energy when combined with \textit{Planck} Cosmic Microwave Background (CMB) \citep[][]{2020A&A...641A...6P} and/or Baryon Acoustic Oscillation \citep[BAO; e.g.,][]{2025JCAP...02..021A} measurements. When combined with these probes (CMB, BAO), current SN samples measure the average dark energy equation-of-state parameter $w$ to $\simeq3$ per cent \citep[]{2018ApJ...859..101S,2022ApJ...938..110B,2024ApJ...973L..14D}. This includes a systematic error budget that is comparable to the statistical uncertainty, and show $w$ to be broadly consistent with a cosmological constant ($w=-1$), with the intriguing possibility of a time-varying $w$ that motivates the need for larger samples of SNe Ia to higher-$z$. Using SNe Ia detected by LSST, TiDES will enable the creation of a Hubble Diagram of ${>}100\,000$ cosmologically-useful SNe Ia and make a ${<}2$ per cent measurement of $w$, when including a CMB prior, and considering only statistical uncertainties. These constraints are a factor 10 improvement compared to DES-SN5YR. Massive spectroscopic follow-up of SNe Ia will also allow their use in different contexts, for example their brightnesses are lensed by structure, tracing matter distributions along their line of sight \citep[][]{1964MNRAS.128..307R,1998ApJ...506L...1H,1999MNRAS.305..746M,2010MNRAS.405..535J,2014ApJ...780...24S,2020MNRAS.496.4051M,2024MNRAS.532..932S,2024MNRAS.tmp.2556S}, and, with LSST, SNe Ia will be used to trace peculiar velocities on large scales and constrain the growth rate of structure \citep[][]{2017ApJ...847..128H}.

The extragalactic time-domain universe is a far more diverse environment than imagined only a decade ago, and exploring the explosive timescale-luminosity phase-space forms the second goal of TiDES. Objects such as: superluminous supernovae \citep[SLSNe; e.g.,][for a review see \citealt{2019ARA&A..57..305G}]{2011Natur.474..487Q,2013ApJ...770..128I,2019MNRAS.487.2215A,2022MNRAS.516.1193K,2023ApJ...943...41C}, calcium-strong (-rich) transients \citep[interchangeably referred to as CaST or CaRT in the literature e.g.,][]{2012ApJ...755..161K,2020ApJ...905...58D}, exotic thermonuclear explosions \citep[see review by][]{2017hsn..book..317T}, fast, blue optical transients \citep[FBOTs; e.g.,][]{2014ApJ...794...23D,2018MNRAS.481..894P,2018ApJ...865L...3P,2019ApJ...872...18M,2023ApJ...949..120H}, tidal disruption events \citep[TDEs; e.g.,][]{1988Natur.333..523R,2012Natur.485..217G,2014MNRAS.445.3263H,2021ApJ...908....4V,2021ARA&A..59...21G}, other ambiguous nuclear transients \citep[][]{2017NatAs...1..865K,2019NatAs...3..242T,2023MNRAS.522.3992W}, and kilonovae \citep[][for a review see \citealt{2021ARA&A..59..155M}]{2013Natur.500..547T,2017Natur.551...64A,2017Sci...358.1556C,2017ApJ...850L...1L,2017Natur.551...75S,2017ApJ...848L..16S,2017ApJ...848L..27T,2017ApJ...848L..24V,2024Natur.626..737L}, have each demonstrated how little is known about explosive transient populations. TiDES will enlarge the population of spectroscopically-confirmed transients by an order of magnitude, and potentially uncover entirely new forms of explosions. This will include ${\sim}$\nCCThree\, core collapse SNe, ${\sim}$\nSLSNThree\, SLSNe for probing the $z>1$ universe, and ${\sim}$\nfaintandFastThree\, faint-and-fast SN Ia-like events for studying the full range of thermonuclear white dwarf explosions.

The third goal of TiDES is cosmology and galaxy evolution with AGN. These are among the most energetic sources in the Universe, showing variability in all wavebands when mass is accreted onto supermassive black holes in the centres of galaxies. The variability of the optical continuum light from the accreting matter, together with its delayed response mirrored in the broad emission lines of the surrounding material can be used i) to dynamically measure black hole masses \citep[e.g.,][]{2016ApJ...818...30S} and identify sub-parsec-scale supermassive black hole binaries among the AGN population, and ii) as a standardisable candle to high redshifts \citep[][]{1982ApJ...255..419B,1993PASP..105..247P,2011A&A...535A..73H,2011ApJ...740L..49W,2023FrASS..1030103P,2023Ap&SS.368....8C}. TiDES will establish an SNe-independent Hubble diagram of AGN and will deliver one of the, if not the, largest catalogue of dynamically measured supermassive black hole masses over redshifts $0.1<z<2.5$. The expected accuracy of both the Hubble diagram as well as the black hole masses will depend critically on the achieved season lengths and total baseline of the survey \citep[][]{2022MNRAS.516.3238M} as well as characterisation of the objects in terms of luminosity and accretion rate \citep[e.g.][]{2016ApJ...825..126D,2020ApJ...899...73F,2023A&A...678A.189P,2024A&A...683A.140Z}.

This paper introduces the TiDES survey and demonstrates its potential using simulations of the LSST and 4MOST survey strategy. In Section~\ref{sec:4mostIntro} we describe the 4MOST spectrograph and the TiDES strategy. Section~\ref{sec:simulations} details our TiDES simulations, and Section~\ref{sec:results} presents sample statistics. Our expectation for the SN Ia cosmological analysis is detailed in Section~\ref{sec:SNIa-cosmology}. We then summarise in Section \ref{sec:summary}. Throughout, we assume a baseline cosmology with a matter density $\Omega_\mathrm{M}=0.315$ \citep[e.g.][]{2020A&A...641A...6P} in a flat universe and a Hubble constant of $H_0=70$\,km\,s$^{-1}$\,Mpc$^{-1}$, and use the AB photometric system \citep{1983ApJ...266..713O}.

\section{4MOST and the TiDES programme}
\label{sec:4mostIntro}

4MOST is a new high-multiplex, wide-field spectroscopic facility under development for VISTA, located on the \lq NTT peak\rq\, at the Cerro Paranal in northern Chile. A full overview can be found in \citet{2019Msngr.175....3D}, with the scientific operations model described in \citet{2019Msngr.175...12W}, and the general survey strategy overview in \citet{2019Msngr.175...17G}. 4MOST has a large field-of-view of 4.2\,deg$^2$ with a high multiplex capability: 1624 fibres feeding two low-resolution spectrographs (LRS), and 812 fibres feeding a high-resolution spectrograph (HRS). The LRS offers $\langle R\rangle=6\,500$ over 3700--9500\AA, and the HRS offers $\langle R\rangle=20\,000$ in three 400--600\AA\ wide passbands. Within a 1-hour observation, 4MOST has the sensitivity to obtain redshifts of $r=22.5$\,mag galaxies or AGN using the LRS. TiDES will use the LRS fibres.

The 4MOST project as a whole consists of 25 consortium and community surveys each with their specific science goals. Each survey is characterised by a target selection algorithm, but all surveys share the focal plane and are observed in parallel, with individual targets for any given observation selected by a fibre-to-target assignment algorithm \citep{2020A&A...635A.101T}. It is expected that TiDES will average around 30--35 LRS fibres in each extragalactic 4MOST pointing -- around two per cent of the LRS fibres available.

\subsection{The TiDES science programme}
\label{sec:tides-science-plan}

The TiDES science programme comprises three interlocking surveys each arising from 4MOST’s ability to conduct extensive spectroscopic follow-up of extragalactic transients located with LSST, whilst exploiting additional diagnostic data from other facilities (e.g., \textit{Euclid}). In this section we describe each of the three surveys.

\subsubsection{TiDES-Live}
\label{sec:tides-sn}

The first survey focuses on the spectroscopic study of live transients (\lq TiDES-Live\rq). This survey aims to observe more than \nSNeAbout\, live transients discovered by Rubin and other photometric surveys as soon as feasible after discovery. There are two main goals:
\begin{enumerate}
    \item A systematic search and classification effort for the full range of extragalactic transients across the luminosity--timescale plane, leading to improved rates and physical understanding of the various events, and 
    \item The rapid follow-up of live SN Ia candidates, which will lead to a new Hubble diagram for cosmological studies, while also serving as a critical spectral training set for the larger LSST SN Ia sample studied (and classified) photometrically.
\end{enumerate}

Work over the last decade has charted the luminosity--timescale plane for transients, and uncovered rare and new classes of transients. TiDES-Live will conduct a systematic campaign of follow-up that will be both unbiased and an order of magnitude larger in number than earlier efforts of this nature \citep[e.g.,][]{2020MNRAS.496...19L}, thereby providing improved rates for both new and well-understood phenomena with full photometric light curves.  Early time spectra ($<$3--4 days post-detection) for all classes of SNe will provide new insights into the progenitor environment and the outer layers of the ejecta, and, in 4MOST pointings with repeat visits, multiple spectra will trace SN evolution into the nebular phase. TiDES will also collect the largest spectroscopic sample of rare transients (e.g., calcium-strong fast transients and TDEs) to comprehensively study their important contributions to galactic chemical evolution.

For SN Ia cosmology, TiDES-Live will create an impressive Hubble diagram of more than 10\,000 cosmologically-useful events over the redshift range $0<z<0.6$ for which there will be associated Rubin deep imaging. A further goal is to explore the dispersion in the Hubble diagram as a function of astrophysical properties (e.g., morphology, SN colour, SN spectral features), with the aim of improving the cosmological precision using SN Ia subsets. Moreover, as even the most advanced machine-learning SN photometric classification techniques are dependent on large, homogeneous and representative training samples \citep[][]{2016ApJS..225...31L,2020MNRAS.491.4277M,2022AJ....163...57Q}, the TiDES sample will become an unsurpassed training sample for SN photometric classifiers \citep{2021MNRAS.508....1C} unlocking the potential of the entire LSST SN dataset.

\subsubsection{TiDES-Hosts}
\label{sec:tides-hosts}

The second survey is spectroscopy of SN host galaxies (\lq TiDES-Hosts\rq). TiDES will secure the spectroscopic redshifts for the host galaxies of SNe of all types for which live SN spectroscopy was not possible, or for both SN and host in regions with dense 4MOST follow up. These redshifts, coupled with photometric classification of the SN events themselves, will form a key component of the LSST SN Ia cosmology programme \citep{2018arXiv180901669T}. 4MOST should secure galaxy redshifts over the range $0<z<1$ in two hours, and to higher redshifts in fields where multiple repeat exposures can be stacked.

In addition to providing galaxy redshifts essential for the LSST SN Ia Hubble Diagram, TiDES-Hosts will provide further diagnostics of the dispersion in SN Ia magnitudes, such as host galaxy metallicity and star-formation rate \citep[following techniques presented in e.g.][]{2023RASTI...2..453D, 2022A&A...659A..89G}. Redshifts also enable the intrinsic properties of individual transients, like luminosity and evolution timescale, to be measured. TiDES-Hosts will provide the largest and most precise measure of the cosmological parameters over the key redshift range where dark energy becomes a dominant component.

In this work we consider host galaxies from TiDES in isolation, but it is important to highlight the contribution of other surveys in 4MOST. There are several participating surveys such as the Wide-Area VISTA Extragalactic Survey \citep[WAVES;][]{2019Msngr.175...46D} and the 4MOST Hemisphere Survey of the Nearby Universe \citep[4HS;][]{2023Msngr.190...46T} that will observe millions of galaxies over thousands-of-square-degrees. Naturally, some of these galaxies will host supernovae, this will benefit TiDES in two ways: i) TiDES will not need to spend fibre-hours re-observing these galaxies, ii) If 4MOST has completed its planned visits of any given field, redshifts will be available from galaxies hosting future supernovae. These additional observations will supplement the total number of available redshifts for any SN analysis.

\subsubsection{TiDES-RM}
\label{sec:into-tides-rm}
The final survey concerns repeat spectroscopic observations of AGN for reverberation mapping (\lq TiDES-RM\rq). The survey will focus on long-term variability monitoring of AGN in the deep drilling fields (DDFs) of LSST \citep{2022ApJS..262...49K,2023A&A...675A.163C}.  Specifically, we will exploit AGN broad line lags to measure dynamical masses of supermassive black holes and construct a Hubble diagram out to redshift $z=2.5$, i.e. into cosmic noon. In addition, the continuous high-quality spectroscopic monitoring will enable us to search for any unusual kinematics of the broad lines that could indicate binary supermassive black holes \citep[for a review see][]{2019NewAR..8601525D}. 4MOST will routinely visit the DDFs every 10-14 days to observe the high density of requested targets and satisfy the deep observing requirements from other surveys. TiDES-RM will use each of the single epoch visits to sample the expected lags of our AGN at a similar rate to past reverberation mapping campaigns \citep[e.g.][]{2011ApJ...743L...4B}. The number of observations required for each object is determined by its lag length, with shorter lags being recovered first. Generally, for a successful recovery of the broad line lags, a single observing season should be longer than a lag length, and an entire observing campaign should be longer than 3 times the lag length.

As standardisable candles, AGN offer a complementary probe to SNe Ia with a reach to higher redshifts and with different astrophysical systematics. Recent work suggests that current probes at low and high redshifts may provide inconsistent results, beyond the well-known $H_0$ tension, pointing either towards unknown systematics in the current methods or new physics \citep[e.g.][]{2019PDU....2600385D,2022MNRAS.516.1721C,2023MNRAS.522.1247K,2024MNRAS.528.6444C,2024ApJ...961..229Z}. TiDES will move forward with a unique test for remaining systematic uncertainties by using two independent standardisable candles, thereby increasing the reliability of the final results. By targeting around 930 AGN over $0.1 < z < 2.5$ we can improve the precision of constraints based on reverberation mapped AGN by at least a factor of two \citep{2015MNRAS.453.1701K}. 

Of equal importance in TiDES-RM will be dynamical mass measures of large numbers of supermassive black holes (SMBH) to $z\sim2.5$, i.e. into cosmic noon, the most active phase of galaxy and black hole growth. Supermassive black hole masses represent a key parameter in governing the assembly history of galaxies through associated feedback processes. Most current SMBH mass measurements outside the local Universe rely on indirect methods, most notably the single-epoch method that takes the width of a broad line and infers the mass via a set of calibrations against local scaling relations.
This method is prone to biases and uncertainties as a single emission line width will be unable to capture the dynamical complexities of the underlying accretion and outflow processes and due to the intrinsic difference in AGN demographics at low and high redshift \citep[e.g.][]{2016MNRAS.460.3119S,2024ApJ...962...67W}. TiDES-RM will measure dynamical SMBH masses at redshifts up to $z \sim 2.5$ via reverberation-mapping, which is the technique underlying the calibration of local scaling relations \citep[e.g.][]{2024ApJS..272...26S}. In addition, the spatio-kinematic information from emission line lags can be used to trace outliers, with the potential of exposing binary supermassive black holes \citep[e.g.][]{2019ApJ...881..140S}.

\section{Simulating LSST+TiDES}
\label{sec:simulations}

\begin{figure*}
    \centering
    \includegraphics[width=0.8\textwidth]{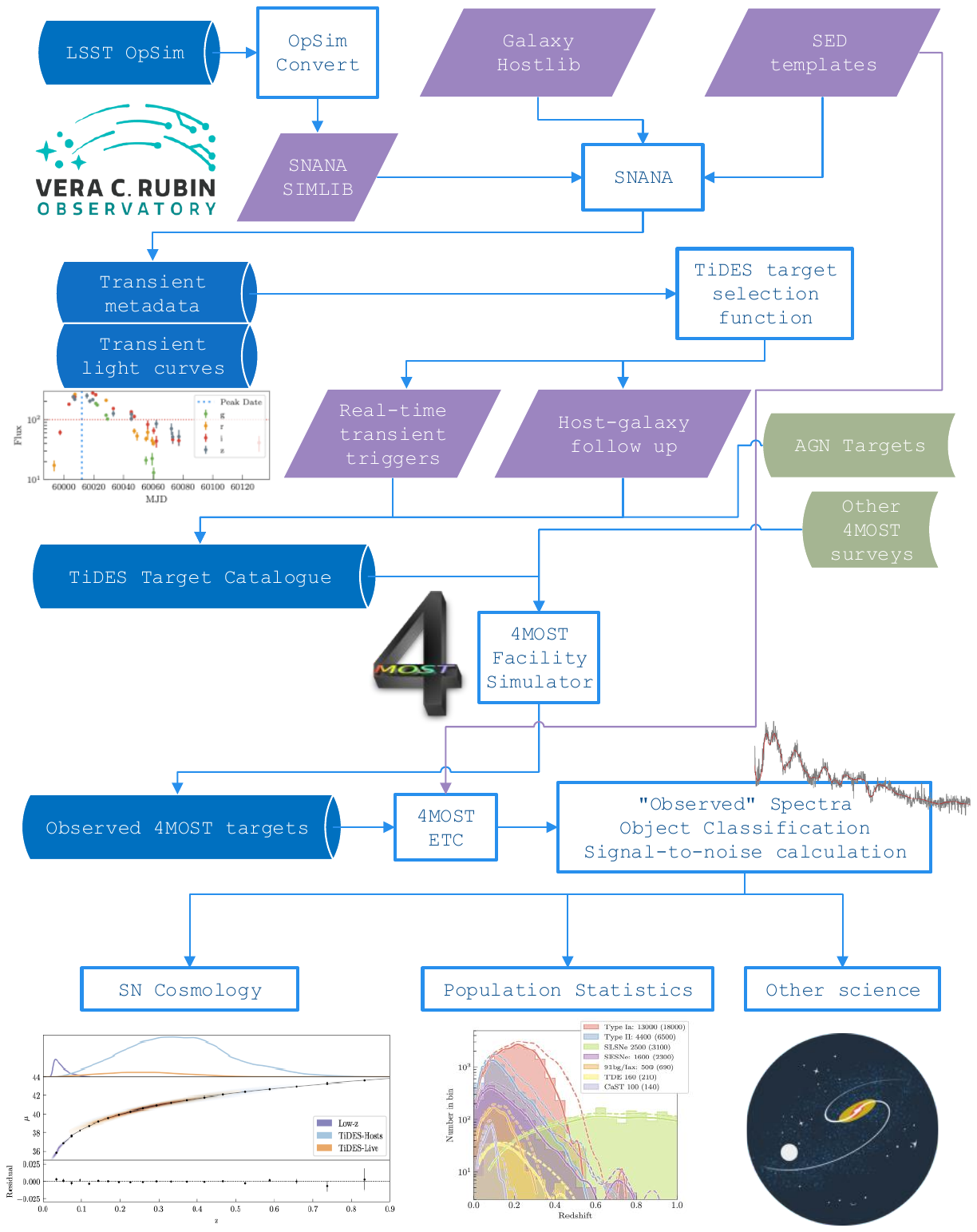}
    \caption{Flowchart of our end-to-end simulation framework. We begin with a database input of a LSST OpSim realisation that is transformed into an \snana\ SIMLIB file. Along with transient spectral templates and a host-galaxy library to draw from, we use \snana\ to create realistic light curves of transient phenomena. The resulting database of events is then queried as if data were flowing in real-time, such that we can trigger for immediate follow up of live-transients or time-insensitive observations of galaxies. These observational requests are merged with our pre-determined AGN targets and ingested into the 4MOST Facility Simulator (4FS) -- along with all other 4MOST member surveys. 4FS allocates fibre-time and scheduling during a mock 5-year survey, returning a description of all observations at the catalogue-level. This data product allows us to generate synthetic spectra with realistic observational effects. These spectra are then assessed for their quality and sent to their appropriate science channels. This allows us to perform a mock cosmology analysis of type Ia supernova cosmology, study our final population statistics, and enables other analyses to be performed.}
    \label{fig:flowchart}
\end{figure*}

We next describe our method for simulating a LSST-like SN survey, and combining with a simulated 4MOST observing strategy. Our simulations focus on LSST for the transient discovery as we anticipate this will provide the majority of the TiDES targets. However, TiDES will be agnostic to the source of the photometric transient feed and will include transients from any source providing a public alert stream e.g. the Zwicky Transient Facility~\citep[ZTF;][]{2019PASP..131a8002B}, the Young Supernova Experiment~\citep[YSE;][]{2021ApJ...908..143J}, BlackGEM~\citep[][]{2022SPIE12182E..1VG}, The Gravitational-wave Optical Transient Observer~\citep[GOTO;][]{2022MNRAS.511.2405S}, and The La Silla Schmidt Southern Survey~\citep[LS4;][]{2025arXiv250314579M}. Thus, this section is intended to produce broad projections for the number and type of transients and hosts that 4MOST may observe, in particular noting that neither LSST nor 4MOST's strategy are yet finalised. An end-to-end flowchart of our entire simulation framework is shown in Fig.~\ref{fig:flowchart}.

\subsection{Simulating transients}
\label{sec:sim_transients}

\begin{table*}
	\caption{Transient classes simulated for LSST using \snana.}
	\centering
\begin{tabular}{p{0.16\linewidth} p{0.32\linewidth} p{0.4\linewidth}}
\hline
\hline
Transient Class & Description & Source \\
\hline
SN Ia & Normal type Ia SNe following SALT2  & \citet{2018ApJ...867...23H} \\
SN Iax & Type Iax peculiar SNe & \citet{2019PASP..131i4501K} \\
SN Ia-91bg & Peculiar SNe SN1991bg-like & \citet{2019PASP..131i4501K} \\
SN II, IIn & Hydrogen-rich core-collapse SNe & \citet{2019MNRAS.489.5802V} \\
SN IIb, Ib, Ic, Ic-BL & Hydrogen-poor stripped envelope SNe & \citet{2019MNRAS.489.5802V} \\
CaRT/CaST & Calcium-rich(-strong) faint and fast SNe  & \citet{2019PASP..131i4501K} \\
SLSN-I & Hydrogen-poor super luminous SNe & \citet{2017ApJ...850...55N,2019PASP..131i4501K} \\
TDE & Tidal disruption events & \citet{2019PASP..131i4501K, 2019ApJ...872..151M}\\
\hline
\label{tab:snClasses}
\end{tabular}
\end{table*} 

The LSST Science Requirements Document \citep{lsstSRD} details the survey specification but with enough room for optimisation of the strategy to balance the different science goals \citep[][]{2022ApJS..258....1B}. LSST will have a ‘Wide, Fast, Deep’ (WFD) component coupled with deeper, more frequent observations in ‘Deep Drilling Fields’ (DDFs). Significant work has been performed on the effects of different survey choices for transients \citep[e.g.,][]{2022ApJS..259...58L,2023ApJS..265...43A,2023ApJS..264...22G}, and we do not attempt to determine an optimal LSST strategy for TiDES in this work. Instead, we use established LSST simulations to create our mock transient datasets, adopting the baseline v3.4\footnote{\href{https://survey-strategy.lsst.io/index.html}{https://survey-strategy.lsst.io/index.html}} simulation, the latest baseline release at the time of writing. The comparisons with other LSST strategies are presented in the TiDES Cadence Note\footnote{ \href{https://docushare.lsst.org/docushare/dsweb/Get/Document-37640/Frohmeier_TiDES.pdf}{https://docushare.lsst.org/docushare/dsweb/Get/Document-37640/Frohmeier\_TiDES.pdf}}. Note that due to the high cadence of repeat visits in the DDFs, all of our TiDES-RM targets are located there.

The output from the LSST Operations Simulator (OpSim)\footnote{\href{https://www.lsst.org/scientists/simulations/opsim}{https://www.lsst.org/scientists/simulations/opsim}} is the input to our simulations.
Based on a scheduling pattern, OpSim creates a catalogue-level output of the LSST strategy using detailed observing quality metrics based on weather models, maintenance requirements, and telescope hardware constraints. Using these observing characteristics, we then simulate transients and their host galaxies using the  SuperNova ANAlysis software \snana\ \citep{2009PASP..121.1028K} simulation code. \snana\ combines survey-specific metadata (e.g., from OpSim) with the known properties of a SN population to generate a set of light curves as observed by any survey strategy. The OpSim strategies are translated into the appropriate \snana\ format using the \textsc{OpSimSummary} package \citep{2020ApJS..247...60B}. Note that all of our simulations are \lq catalogue level\rq, i.e., we do not place transients and hosts into simulated images. We assume a saturation limit of 15 mag for the LSST CCD and exclude all points on a simulated light curve exceeding this brightness.

\snana\ generates light curves by integrating a time-evolving spectral energy distribution (\lq template\rq) of a given SN through the survey’s filter-specific response function. Table~\ref{tab:snClasses} shows each spectral template we use in our simulations. Our method of generating light curves from spectral templates using \snana\ follows the Photometric LSST Astronomical Time Series Classification Challenge (PLAsTiCC)  \citep{2019PASP..131i4501K}, but we update the core collapse SN templates to those of \citet[][hereafter V19]{2019MNRAS.489.5802V}.

On a technical note, \snana\ precomputes a set of transient sky coordinates with a resolution of several arcminutes. This leads to some simulated events sharing spatial coordinates, which if left unaddressed would lead to artificial fibre collisions in the 4MOST simulator. Increasing this resolution in \snana\ becomes computationally expensive, so we perform a post-processing stage to apply a small coordinate scatter on each transient and host such that their distribution is uniform within an \snana\, resolution element.

\subsubsection{Simulation of SNe Ia}
\label{sec:simSNeIa}

We simulate SNe Ia using the SALT2 SED model \citep{2007A&A...466...11G, 2010A&A...523A...7G} trained following \citet{2014A&A...568A..22B}. The SALT2 SED model is described by the SN redshift $z$, the epoch of SN peak brightness $t_{\mathrm{peak}}$, a (dimensionless) stretch-like parameter $x_1$, and a colour parameter $c$. The SN Ia apparent magnitude in the rest-frame $B$-band $m_B$ is then simulated as
\begin{equation}
\begin{split}
    m_B = &- \alpha x_1 + \beta c +\mathcal{M}_B\\ 
    & + 5\log_{10}\left ( d_L\left (z, \Omega,w\right )\right / 10\, \mathrm{pc} ) + \delta m_B,
\end{split}
    \label{Ia_std}
\end{equation}
where the parameters $\alpha$ and $\beta$ describe the stretch--luminosity and colour--luminosity relations and are set in our simulations to $\alpha=0.14$, and $\beta=3.1$. The absolute magnitude for a $x_1=0$, $c=0$ SN Ia ($\mathcal{M}_B$) is set to -19.365 and $d_L$ is the luminosity distance for redshift $z$ with our default cosmological parameters. The term $\delta m_B$ is a randomised brightness offset that encapsulates the intrinsic scatter observed in SN Ia standardised brightness \citep[see][for details on the modelling of intrinsic scatter]{2010A&A...523A...7G}. The absolute number of SN Ia events simulated and their redshift distribution is calculated using the power-law evolution of the volumetric rates presented by \citet{2019MNRAS.486.2308F} (their Equation 6). The intrinsic distributions of SN stretch $x_1$ and colour $c$ are taken from \citet{2016ApJ...822L..35S}.

\subsubsection{Simulation of other types of transients}
All simulated transient types are listed in Table~\ref{tab:snClasses}. For core-collapse SNe, we use \citetalias{2019MNRAS.489.5802V} templates and include both hydrogen-rich core-collapse SN events (type II and IIn SNe) and stripped envelope SNe (Type Ib/IIb/Ic/Ic-BL SNe). For volumetric rates of core-collapse events, we normalise at low redshift to the measurement of \citet{2021MNRAS.500.5142F} and extrapolate to higher redshifts assuming the rate follows the star formation history presented in \citet{2014ARA&A..52..415M}. To ensure the correct mix of different core collapse sub-types we use the relative rates presented by \citet{2017PASP..129e4201S}. The intrinsic brightness of simulated events are matched to the luminosity functions presented in \citet{2011MNRAS.412.1441L} and revised by \citetalias{2019MNRAS.489.5802V}. Recently, updates have been made to models using different rate assumptions for Type IIn SNe \citep[][]{2025ApJ...987...13R} and templates for stripped-envelope \citep[][]{2024ApJS..275...37K}, however, these updates do not affect the overall picture for TiDES compared to the models we have used.

We include two types of peculiar SNe Ia: SN1991bg-like events \citep[][]{1992AJ....104.1543F} and SNe Iax \citep[][]{2013ApJ...767...57F}. SN1991T-like events are modelled within the normal-SN Ia events in Section~\ref{sec:simSNeIa}. The SED templates and volumetric rates used to simulate these classes of peculiar SNe Ia are those presented by \citet{2019PASP..131i4501K}, applying the revisions in \citet[][hereafter V21]{2021MNRAS.505.2819V}.

Tidal Disruption Events (TDEs), Calcium-Strong Transients (CaST) and hydrogen-poor Superluminous SNe (SLSNe-I) are implemented following the same assumptions and modelling used in the PLASTICC challenge \citep{2019PASP..131i4501K}.

\subsubsection{Simulation of AGN for TiDES-RM}
\label{sec:tides-rm}

The AGN that we will observe as part of the reverberation mapping survey are already known and are located only in the LSST DDFs. The selection is based on two catalogues, the Million Quasar Catalogue Milliquas \citep[][]{2023OJAp....6E..49F} and the southern photometric quasar catalogue based on Dark Energy Survey DR2 as presented by \citet{2023ApJS..264....9Y}. First, both catalogues were cut down to the hexagonal footprints of the 4MOST field-of-view centered on the defined 4MOST deep field positions within the Elais South, COSMOS, ECDF-S, and XMM-LSS deep fields (Table~\ref{tab:agnsamp}). For Milliquas, only sources with a \textit{Q} or \textit{A} type flag have been included, which selects against obscured AGN that are unfavourable to reverberation mapping. In a second step, sources from both catalogues were cross-matched. Those present in both catalogues are deemed the ``gold sample'' while objects only present in Milliquas are part of the ``silver sample.'' The total number of identified AGN varies with deep field, given how intensely the respective field has been monitored and in which wavebands. Given survey operational constraints, the maximum number of TiDES-RM sources that can be observed per deep field is of the order of 250. To preserve some comparability across the fields, we impose brightness cuts of $r < 20.5$ for the COSMOS and XMM-LSS fields and $r < 21.5$ for the Elais South and ECDF-S fields. 



Finally, we want to exclude sources where the survey length will not be able to recover a lag. This requires estimating observed lag length of the cut down catalogues. We use the lag-luminosity relation for the H$\beta$ line presented in \citet{2009ApJ...697..160B} and scale to \ion{Mg}{2} lags with a factor of 0.9. \ion{C}{4} lags are estimated from the lag-luminosity relation in \citet{2022MNRAS.509.4008P}. Approximate luminosities are calculated from $r$ band magnitudes. The resulting estimated rest-frame lags have been converted to observed lags by accounting for a cosmological time dilation factor of $(1+z)$. We assume that successful lag recovery requires lag lengths to be at least 3 times shorter than the survey length. As such, we cut the target catalogue for AGN where the shortest lag observed in the 4MOST wavelength range for the given redshift is $>$600 days. With the previously imposed brightness limit, this primarily affects the brightest sources in the redshift range $0.6 < z < 1.6$ and particularly those at the higher redshift end, where \ion{Mg}{2} is the only one of the 3 main target lines present. For all four DDFs, this affects $\le$1\% of the pre-cut samples.

\begin{table*}
\caption{The centres of the 4 DDFs that 4MOST will routinely observe during the 5-year survey. Within our TiDES-RM program (Section~\ref{sec:tides-rm}), these fields will be visited every $14\pm4$ days. Number of AGN observed in each DDF, including sample distribution across three prominent broad lines that will be monitored.} \label{tab:agnsamp}
    \begin{center}
    \begin{tabular}{l c c c c c c}
    \hline\hline
    Field & R.A. & Dec. & AGN & H$\beta$ & \ion{Mg}{2} & \ion{C}{4} \\ \hline
    ELAIS-S1 & 9.500   & -43.950 & 	191 &	23 &	128 &	80 \\
    XMM-LSS & 35.875  & -5.025 &	252 &	47 & 	141 &	119 \\
    ECDF-S  & 53.125  & -28.100 &	230 &	48 &	138 &	92 \\
    COSMOS & 150.125 & 2.200 &	256 &	47 &	150 &	94 \\ \hline
    \end{tabular}
    \end{center}
    \textit{Notes} --- Redshift ranges covered by individual lines are: H$\beta$: $z < 0.64$; \ion{Mg}{2}: $0.43 < z < 1.85$; \ion{C}{4}: $1.58 < z < 4.16$. Spectra of some sources will include two broad lines that can be used to cross-calibrate lags: ELAIS-S1: H$\beta$/\ion{Mg}{2}: 10, \ion{Mg}{2}/\ion{C}{4}: 30; XMM-LSS: H$\beta$/\ion{Mg}{2}: 17, \ion{Mg}{2}/\ion{C}{4}: 38; ECDF-S: H$\beta$/\ion{Mg}{2}: 20, \ion{Mg}{2}/\ion{C}{4}: 28; COSMOS: H$\beta$/\ion{Mg}{2}: 14, \ion{Mg}{2}/\ion{C}{4}: 21.
\end{table*}

The final sample numbers across the four DDFs are listed in Table~\ref{tab:agnsamp}. We highlight number of objects per emission line as well as number of sources in those redshift regions where two lines overlap. These are particularly interesting as they will allow for cross calibration of lag-luminosity relations as well as black hole mass estimates. In total, 61 AGN are expected to have both H$\beta$ and \ion{Mg}{2} lags recovered and 117 to overlap in \ion{Mg}{2} and \ion{C}{4}.

\subsection{Simulating transient host galaxies}
\label{sec:sim-hosts}
SNe Ia, peculiar SNe Ia and core-collapse SNe are associated to host galaxies following the approach and assumptions presented in \citetalias[]{2021MNRAS.505.2819V}. Host galaxies are selected from a \lq HOSTLIB\rq\ library of 380\,000 galaxies from the DES Science Verification galaxy catalogues \citep{2018ApJS..235...33D}. Each galaxy in the catalogue has optical photometry from DES ($griz$) and -- when available -- near infrared photometry from VISTA. To derive properties of the galaxies, template SEDs were generated using the \textsc{P\'EGASE.2} spectral synthesis code and fit to the observed galaxy fluxes through a $\chi^2$ minimisation; this routine is extensively described in \citet{2010MNRAS.406..782S} and, in the context of DES galaxy catalogues, \citet{2020MNRAS.494.4426S}. From the best fitting templates, we adopt the associated parameters such as: redshift, stellar mass, and star formation rates throughout the rest of this work. The catalogue is complete to $\simeq24.5$\,mag in $r$-band, which is well-matched to the transient hosts that LSST should detect \citep[e.g.,][]{2020MNRAS.495.4040W}.

Host galaxies are associated to simulated SNe Ia following the observed rates as a function of the host-galaxy properties as measured by \citet{2006ApJ...648..868S}. Within each galaxy, we follow the standard \snana\, prescription of placing transients in the host following a S\'ersic profile \citep[][]{1963BAAA....6...41S}. We also ensure that the well-known correlation between the SN Ia $x_1$ parameter and host galaxy stellar mass is reproduced in the simulations (see section 4.5.1 in \citetalias{2021MNRAS.505.2819V} for details), but we do not include the more subtle relation between SN Ia Hubble residuals and host galaxy properties, such as stellar mass, colour, and specific star formation rate.

For peculiar SNe Ia, we scale the SN Ia rate and assign host galaxies so that SN1991bg-like events are only associated to old stellar environments \citep[e.g.][]{2011MNRAS.412.1441L,2019PASA...36...31P} and SN Iax events are placed in star forming galaxies \citep[e.g.][]{2013MNRAS.434..527L,2018MNRAS.473.1359L,2020MNRAS.493..986T}. For core-collapse SNe, we use the rates and host galaxy association measurements presented by \citet{2017ApJ...837..120G,2017ApJ...837..121G}. For the remaining transients their relation to host galaxies is not as extensively studied due to their rarer intrinsic and observational rates; we assign them to hosts randomly and position them within the host following a S\'ersic profile. This random host association for the sub-dominant populations has no significant effect on the TiDES host galaxy statistics.

\subsection{Simulating AGN}

AGN simulations have been carried out based on properties of the selected sample. Due to the nature of optical selection in the target catalogues, the overwhelming majority of the sources will be broad line AGN, as desired for a reverberation mapping experiment. We identify each selected source with a spectroscopic template of a type 1 AGN as provided by the 4MOST AGN survey \citep{2019Msngr.175...42M}. The associated template spectrum for each object has been redshifted and flux scaled according to the catalogue $r$-band magnitude and fed into the 4MOST operations simulator. Within a 4MOST simulation, the AGN templates are used to determine the required exposure times in the DDFs to meet the spectral success criteria (Section~\ref{sec:ssc}).

\subsection{Simulating 4MOST}
\label{sec:4most-strat-sim}

\begin{figure*}
    \centering
    \includegraphics[width=\textwidth]{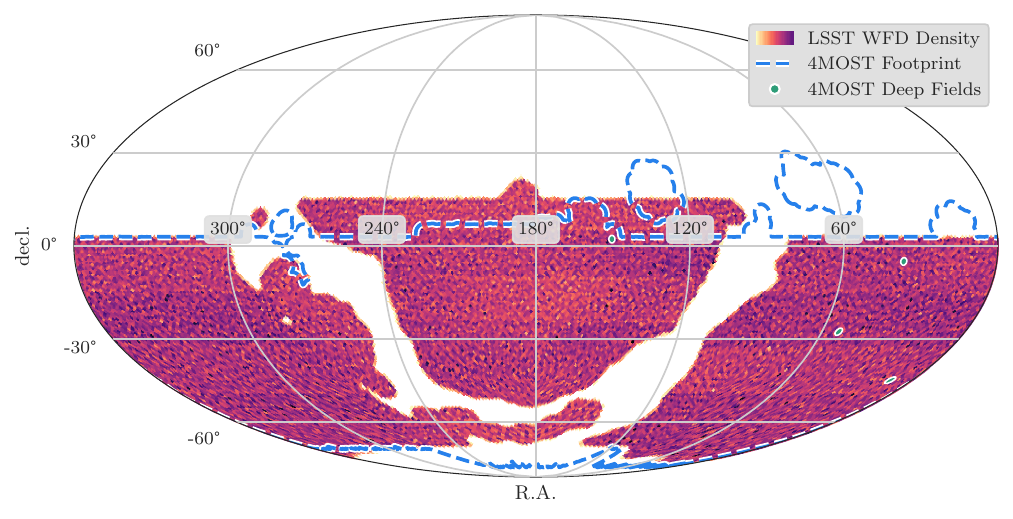}
    \caption{The overlap in footprints of the 4MOST and LSST surveys. The heatmap represents the number of visits in the wide-fast-deep LSST program under the \texttt{baseline v3.4} survey strategy during the overlap with 4MOST operations. Due to the non-uniform sky coverage of LSST over this time period, relics of the rolling cadence seasons can be see as density difference along declination stripes. The dashed blue lines represents the extent of the 4MOST footprint. Within this footprint, certain areas such as the WAVES fields \citep{2019Msngr.175...46D} or LMC/SMC observations \citep{2019Msngr.175...54C}, are observed more frequently to meet the science requirements of the other 4MOST consortium surveys. The green circles show the location of the four DDFs in 4MOST that coincide with the LSST DDFs. 4MOST will observe these locations at a higher cadence -- at least every 14\,d for the TiDES-RM experiment. }
    \label{fig:skyResults}
\end{figure*}

Having generated a simulation of transient light curves as observed by LSST, we next determine which are available to be observed by 4MOST. The 4MOST observing strategy fulfils a complex balance of science requirements from multiple different surveys focused on both galactic and extra-galactic fields. To accommodate this, 4MOST will tile the entire southern-sky over the course of its five-year lifetime \citep[][]{2019Msngr.175....3D}, with the density of targets and exposure time requirements ultimately dictating the frequency of any returning visits. The overlap between the LSST wide-fast-deep footprint and the 4MOST union of observing tiles is shown in Fig.~\ref{fig:skyResults}.

There are two important considerations in the apparent overlap between LSST and 4MOST. The first is that only a small fraction of live LSST transients will ever be observed by 4MOST due to the significant difference in the nightly sky coverage between the two surveys. The second is that although TiDES does have priority fibre-allocation on transient targets and their host galaxies in any field that 4MOST observes, it does not (and cannot) operate in a \lq Target of Opportunity\rq\ mode and force observations of the most interesting targets by changing the telescope pointing. The exact pointing of 4MOST is not known more than a few days in advance, and can change according to predicted or actual observing conditions. Thus, at any one time, all possible LSST transient targets need to be available for insertion into the 4MOST queue, depending on the exact observing pointings that are chosen on a given night.

4MOST has experimented with several mock strategies to establish its observing plan. The success of any strategy is evaluated using a figure-of-merit for each 4MOST survey, combined into a global figure of merit. We evaluate the success of TiDES using the total number of objects observed and the quality of their resulting spectra. Our simulations in this paper use a typical 4MOST mock strategy that has been optimised to a solution that tiles the sky with observations \citep{2020MNRAS.497.4626T} using probabilistic fibre assignments \citep{2020A&A...635A.101T}. From the TiDES perspective, we can use these simulations to track  which of our simulated transients were observed and the exposure time on that epoch. For the host galaxies, we are able to assess the total integrated time the target received over multiple epochs.

\subsection{LSST-TiDES selection function}
\label{sec:selectFunc}

We next describe the process of selecting transients from the LSST simulations for 4MOST spectroscopic follow-up. The aim is to reflect the \lq real-life\rq\ operation of TiDES, where live SNe are identified through difference imaging and distributed in alert packets in the LSST data stream, we will identify suitable targets through custom filtering in community full-stream alert brokers e.g. Lasair~\citep[][]{2019RNAAS...3...26S}, ALeRCE~\citep[][]{2021AJ....161..242F}, ANTARES~\citep[][]{Matheson_2021}, AMPEL~\citep[][]{2019A&A...631A.147N}, FINK~\citep[][]{2021MNRAS.501.3272M}, Pitt-Google~\citep{2024lsst.confE..13W}, and Babamul. This means that new targets must be frequently added to the TiDES target list, and targets that have faded must be removed.

The depth of LSST is fainter than the spectroscopic limiting magnitude of 4MOST, allowing us to track the rising brightness of a transient before submitting it to the 4MOST observing queue -- resulting in a high purity of candidates. Significant advances have been made in rapid real-time classification of transients and spectroscopic follow-up recommendation algorithms \citep[e.g.,][]{2019PASP..131k8002M,2020arXiv201005941K,2022AJ....163...57Q,2023ApJ...954....6G,2024MNRAS.533.2073M, 2024MNRAS.531.2474S}. However, TiDES has access to sufficient fibres to observe all live transients\footnote{In well-sampled LSST fields, ${\sim}12$ live-transients in a 4MOST FoV pass our selection criteria} in any field down to its limiting magnitude, and so is able to adopt a very straight forward selection of potential targets. This also has the advantage of a simpler calculation of the efficiency and selection function for any subsequent TiDES population analysis, including for SN Ia cosmology.

Our planned selection criteria for LSST transients, and those that we use in these simulations, are:
\begin{enumerate}
    \item The transient must be detected to $>5\sigma$ in at least three bands over at least two nights,
    \item The object must be $\le22.5$\,mag in any $griz$ filter (Section \ref{sec:live-sn-pop} for justification).
\end{enumerate}
We only consider $griz$ data points, regardless of additional observations in $uY$ of the field. Clearly, different filter selection criteria would be adopted for transients that did not originate from the LSST alert stream. 

Once all criteria are met, the target is added to a mock TiDES observing queue with a nominal lifetime of four days, after which the target is retired. Each object, however, is monitored so that each subsequent LSST observation of the target that is brighter than 22.5\,mag refreshes the target in the queue by extending its expiry date by an additional four days. This process ensures a high probability that any target selected by the 4MOST scheduler will be bright enough to have a successful spectroscopic observation (see our Spectral Success Criteria (SSC) in Section~\ref{sec:ssc}).

The requirements for the TiDES-Hosts programme are similar to that of TiDES-Live, but with no magnitude limit on the SN. We replace this with a requirement that the (now faded) SN light curve must have shown an increase in brightness followed by a decrease in brightness in at least two filters, followed by at least 50 days undetected in LSST data. This acts as a loose proxy for estimating when the transient peaked in brightness and faded below a detectable limit. We use this requirement as it ensures that the SNe in the selected hosts had both the pre- and post- maximum light data required for a reliable light-curve fit and photometric classification. This is needed for one of our primary science goals, constructing a SN Ia cosmology sample using the host-galaxy redshift. Additionally, we require that the host galaxy has $r\leq\,24$, as targets fainter than this will never meet our SSC. Finally, TiDES-Hosts targets have no expiry date in the 4MOST queue. 

It is possible to imagine different selection criteria and have several streams operating in parallel, each optimised towards transients with differing properties. Adapting some of our requirements may, for example, enable earlier triggering of young or rapid transients at the risk of increasing contamination from spurious events. The selection criteria presented in this work are the result of a conscious effort to maximise the quality of the targets we observe given our available fibre hours, but these may be revised once LSST and 4MOST are operational. 

In the next section, we present the simulations, generate realistic spectra based on the observational metadata, and evaluate the expected samples.

\section{Expected results from 4MOST/TiDES}
\label{sec:results}

The results from our simulations provide us with a catalogue of SNe, their hosts, and AGN that were \lq observed\rq\ as part of the survey realisation. This includes metadata for the observational conditions, including the seeing, exposure time, and airmass. Combining all of this, we are able to generate mock spectra for each of our observed events, evaluate the quality of the mock spectra, and then quantify the performance of the survey simulation. We note that the conclusions we draw here are demonstrated at the population-level, and we do not consider any individual object assessment or seek to derive astrophysical properties from the spectra. Based on the density of targets and signal-to-noise requirements for each of our three science programmes, our fibre-hours are proportionally spent ${\sim}10\%, {\sim}75\%, \mathrm{and}\, {\sim}15\%$ between TiDES-SN, TiDES-Hosts, and TiDES-RM respectively.

\subsection{Mock spectra}
\label{sec:mockSpectra}

To generate mock spectra, template spectra for each observed event are used as inputs to the 4MOST exposure-time calculator (ETC)\footnote{\url{https://escience.aip.de/readthedocs/OpSys/etc/master/index.html}}, with parameters matching the observing conditions from the simulations. For the TiDES-Live survey, we generate phase-specific spectra based on the date of the 4MOST observation using the same templates described in \ref{sec:sim_transients}. Where the same SN is observed more than once, we do not stack the spectra to improve the signal-to-noise due to evolution in the target spectrum. Conversely, for the TiDES-Hosts our targets can be repeatedly observed on multiple occasions and are not expected to evolve, and we take the total integrated exposure time over the survey when generating our final spectra. The ETC returns the instrument-level signal, noise, sky background, and the efficiencies for each of the three arms on the LRS. 

\begin{figure*}
    \centering
    \includegraphics[width=\linewidth]{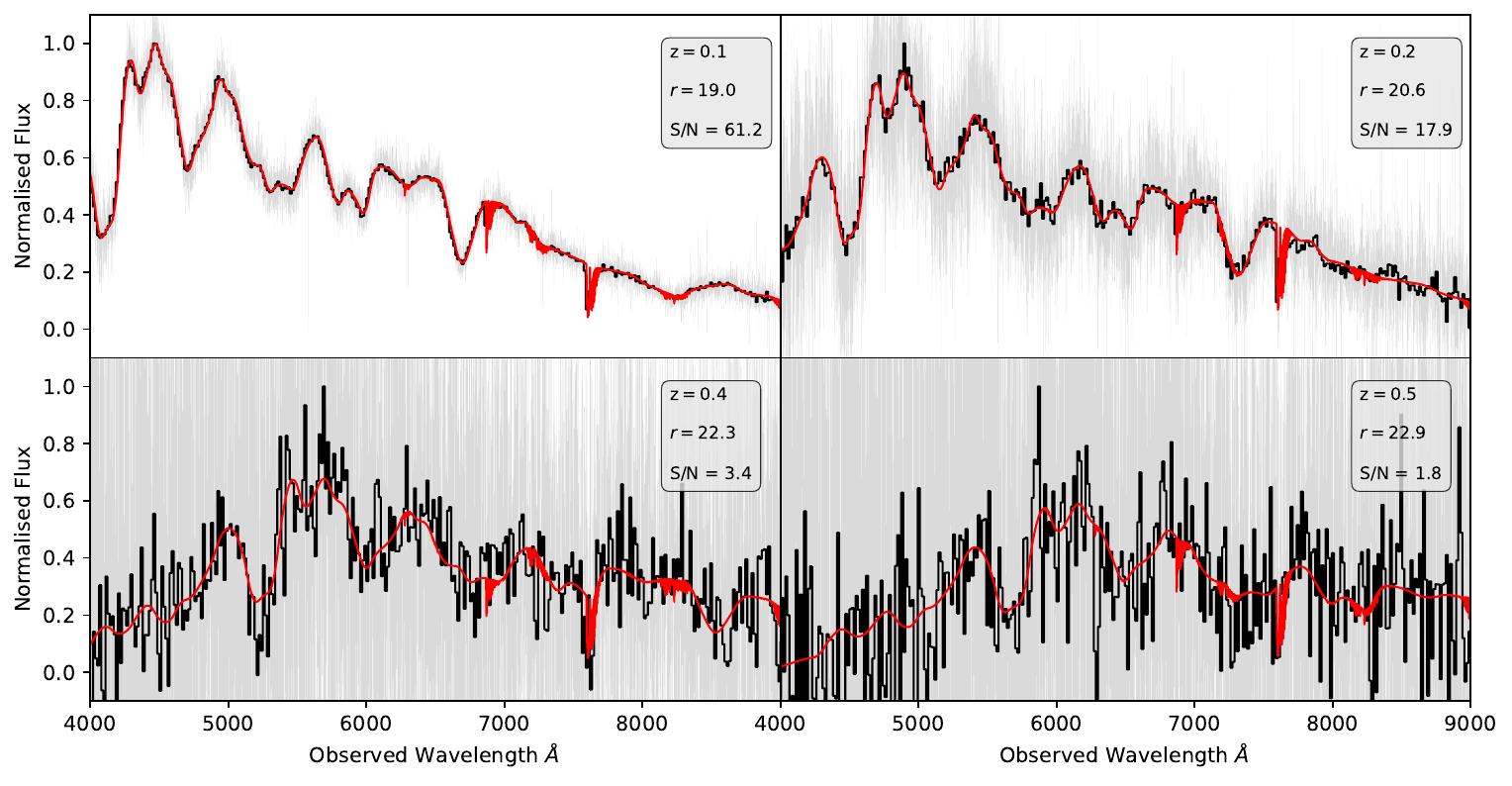}
    \caption{Mock 4MOST spectra based on the 4MOST ETC output. A typical maximum-light SN Ia template spectrum at several different redshifts (red line) and $r$-band magnitudes is shown. The template is \lq observed\rq\ by 4MOST with the raw data shown in light grey and then rebinned to 15Å (black line). The mock spectra are calculated assuming a typical 4MOST exposure time of 2700s, conditions of seeing of 0.8”, airmass 1.4, in grey time. The mean S/N per 15\AA\ bin over 4500-8000\AA\ is also given.}
    \label{fig:exampleSpectra}
\end{figure*}

An example spectrum for a peak-phase SN Ia, observed at different $r$-band magnitudes, is shown in Fig.~\ref{fig:exampleSpectra}. The spectra for TiDES-Hosts (and TiDES-RM) are generated in an similar way.

\subsection{Spectral success criteria}
\label{sec:ssc}

Once our targets have been simulated and observed by 4MOST, we use \lq Spectral Success Criteria\rq\ (SSC) to determine whether an observation was successful. Each of the three surveys in TiDES has its own SSC. By design, objects that reach their SSC are not allocated any further fibre-hours on repeat visits to that field. This constraint is only relevant to the TiDES-Hosts survey as our other programs require the SSC to be achieved in a single visit.

For TiDES-Live, our actual success will depend on the signal-to-noise ratio (SNR) in the 4MOST spectrum and the transient type, due to the prominence of certain spectral features that define the different classes. SN spectra are typically dominated by broad features in the photospheric phases that are many tens-of-Angstroms wide, and thus our SSC are defined using the mean SNR in 15\AA\ bins -- we denote this with the \snrF\ parameter. Our criteria are based on earlier studies of high-redshift SNe Ia \citep{2009A&A...507...85B}, where a robust classification was achieved with a mean \snrF\,=\, 5 over 4500--8000\AA\ in the observer frame. This is a conservative criterion as \citet{2009A&A...507...85B} also demonstrate probable classifications of transients with a mean \snrF\,=\, 3.

We have checked our likely classification limits for SNe Ia using the ETC based on simulated mock spectra with classifications based upon the machine-learning SN classification tool \textsc{dash} \citep{2019ApJ...885...85M}. \textsc{dash} can classify, with 95 per cent confidence, a SN Ia spectrum with \snrF\,=\, 5 without using the host redshift as a prior. These simulations were performed with SN spectra free of host galaxy contamination (e.g., the SN Ia spectra in Fig.~\ref{fig:exampleSpectra}). An investigation into the effect of host galaxy contamination on SN classification in 4MOST will be presented in Milligan et al.(in prep).

For TiDES-Hosts, we require a mean SNR$>$3 per \AA\  over 4500--8000\AA\ (\snrO) in the observer-frame. Galaxies with strong emission lines in real observations can have a redshift successfully measured before reaching this SNR criterion. In our simulation, however, we only obtain the final spectrum SNR rather than any intermediate stages that would have contributed to the stacked spectrum, we assume every galaxy reaching our SSC has had a redshift sucessfully measured.

For TiDES-RM, spectral success is defined by achieving a \snrF=10 for an AGN spectrum. Just as for the TiDES-Live program, the spectra vary with time and, therefore, cannot be stacked if observations are taken on significantly different epochs. Within 4MOST, the other participating surveys have placed a high demand of time on the LSST DDFs. This necessitates many repeat visits that TiDES will exploit to routinely observe our AGN for a reverberation mapping experiment. To meet the SSC, the TiDES-RM targets must reach \snrF=10 in a single visit every ${\sim}14$\,d.

\subsection{Live supernova populations}
\label{sec:live-sn-pop}
Combining the mock spectra for all our targets and the SSC, we are able to construct an expected TiDES-Live sample. In many cases, we split our population statistics into groups for objects satisfying the \snrF\,${\ge}5$ and ${\ge}3$ criteria. As expected, LSST's WFD survey program provides the bulk of our transients, ${\sim}95$ per cent, due to the much larger footprint compared to the DDFs.

\begin{figure}
    \centering
    \includegraphics[width=\linewidth]{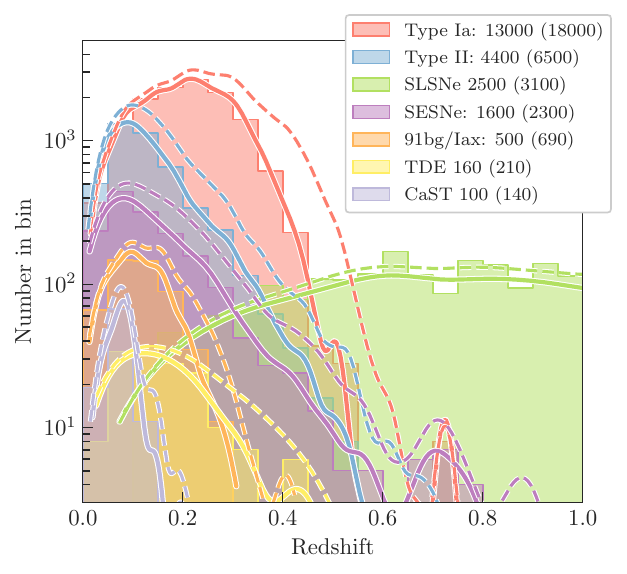}
    \caption{The observed 4MOST redshift distribution for all simulated transient types after five years of 4MOST operations. The solid lines show the expected number of transients with a \snrF\,${\ge}5$, while the dashed lines show the expectation for a \snrF\,${\ge3}$ in the spectrum's continuum. The legend shows the total number of objects for each subclass satisfying the SSC. Different science cases will require different qualities of data, so these numbers are presented as a guide to the final sample size.}
    \label{fig:nz_SNR5}
\end{figure}

We begin with the most general overview metric for any wide-field sky-survey -- the redshift distribution. Fig.~\ref{fig:nz_SNR5} shows the redshift distribution split into the different transient sub-samples from our simulations. Our results show that we can expect to observe around \nSNeAbout\, SNe with (\snrF${\ge}3$) in total. Due to their intrinsic brightness, SNe Ia are the most common transient that TiDES will observe, totalling around \nIasnrThree\ objects with a peak in the redshift distribution at z${\sim}0.3$. This results in the largest homogeneously collected SN Ia sample from any spectroscopic survey. As expected, the SN1991bg-like and SN Iax under-luminous populations of SNe Ia are found at lower redshifts and are intrinsically rarer than the normal counterparts.

SNe II are intrinsically the most common of all SN sub-types \citep[e.g.,][]{2011MNRAS.412.1441L,2017ApJ...837..121G} and this is reflected in the first two redshift bins of the histogram in Fig.~\ref{fig:nz_SNR5}. However, their lower luminosity results in a lower peak redshift for the sample at z${\sim}0.15$. Stripped-envelope SNe (SESNe) also belong to the massive-star progenitor, core-collapse SN family, but occur at $\sim24$ per cent of the intrinsic rate of SNe II \citep[][]{2021MNRAS.500.5142F}. In our sample, SESNe follow a similar observed trend with a peak redshift of z${\sim}0.15$.

SLSNe-I form a class of the brightest observed SNe with a luminosity function ranging from $-19 < \mathrm{M} < -23$   \citep[][]{2019MNRAS.487.2215A}. As a result, they are observable to the highest redshifts of any objects in our simulated sample, and their intrinsically long (and time dilated) light curves increases the opportunities for TiDES to obtain a spectrum. We have truncated the SLSN redshift distribution in Fig.~\ref{fig:nz_SNR5} for visualisation purposes, but note that we expect to obtain SLSNe-I up to z${\sim}2$.

Given the intrinsic luminosity, rates, and light curve evolution timescales of both Calcium-strong transients (CaST) and TDEs, it is unsurprising that these objects form the smallest populations of our samples. Furthermore, as these are some of the least well understood transient classes, any recent updates to their observed properties will not be reflected in our transient models from PLAsTiCC. For example, a slightly brighter TDE luminosity function than the one we adopt is presented in \citet[][]{2023ApJ...955L...6Y}. Regardless, the TiDES sample of these transients will be a significant increase over current sample sizes.   

\begin{figure}
    \centering
    \includegraphics[width=\columnwidth]{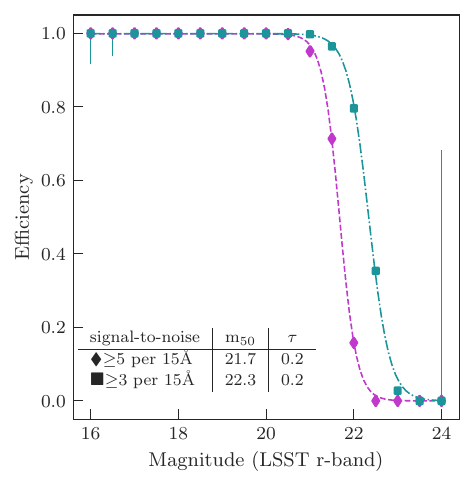}
    \caption{The spectroscopic efficiency of targeted objects as a function of their observed magnitude. Our measurements are shown for objects meeting the \snrF\, $\ge5$ (purple diamonds) and $\ge3$ (green squares) parameter. $1\sigma$ statistical-only uncertainties on efficiencies are calculated using Bayes Theorem following the method presented in \citet[][]{paterno04} and, in the context of supernova survey efficiencies, \citet[][]{2017ApJS..230....4F}. The table shows the best-fit parameters for Equation~\ref{eqn:detEff}, $m_{50}$ is the magnitude corresponding to a recovered fraction of 50 per cent, and $\tau$ describes the steepness of the exponential fall off.}
    \label{fig:effFuncMag}
\end{figure}

We analyse the efficiency of our survey as a function of the target object's magnitude in Fig.~\ref{fig:effFuncMag}. We define the efficiency as the fraction of objects meeting the SSC once a fibre has been placed on them. To estimate these efficiencies, we fit the following sigmoid of the functional form \citep[][]{2007ApJ...660.1165S,2014AJ....148...13R}:
\begin{equation}
    \varepsilon(m) = \eta_{0}\left ( 1 + \exp \left ( \frac{m - m_{50}}{\tau} \right ) \right )^{-1},
    \label{eqn:detEff}
\end{equation}
where $\varepsilon$ is the efficiency, $\eta_{0}$ is the maximum efficiency, $m$ is the magnitude of the object in the LSST $r$-band at the time of observation, $m_{50}$ is the magnitude at which 50 per cent of the sample meet our SSC, and $\tau$ captures the exponential roll-off (smaller numbers describe a steeper drop).

For the \snrF\,$\ge5$ parameter, we find $m_{50}=21.7$ and for \snrF\,$\ge3$, $m_{50}=22.3$. In both cases, the maximum efficiency is $\eta_{0}=1$ and $\tau=0.2$. Live transient objects are added to, or removed from, the 4MOST follow-up queue based, in part, on their observed LSST magnitudes. Using the $m_{50}=22.5$ result as our follow-up limit ensures that we minimise observations resulting in spectra with a SNR too low for a classification.

\subsection{Observational efficiencies of type Ia supernovae}
\label{sec:sneIastats}

\begin{figure*}
    \centering
    \includegraphics[width=\linewidth]{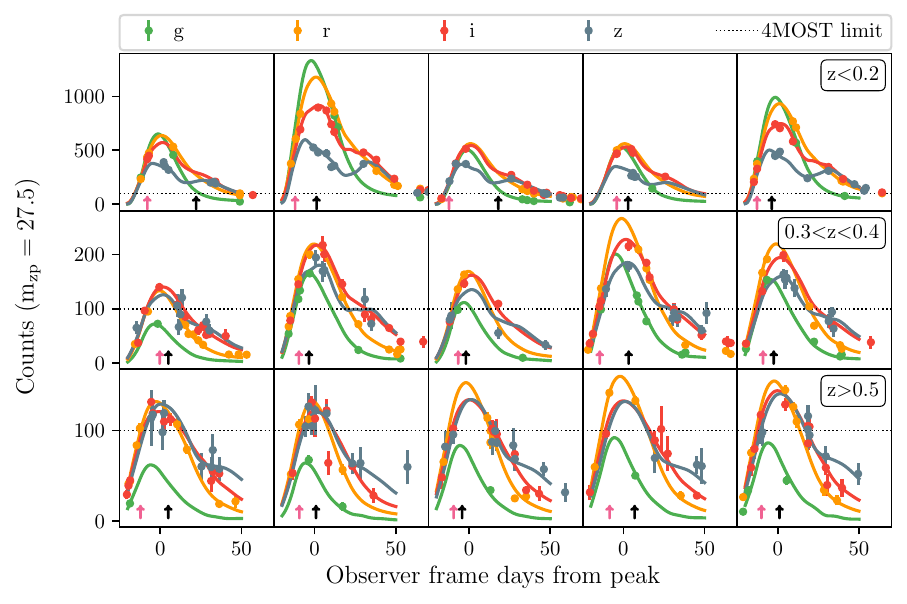}
    \caption{Example observer-frame light curves of SNe Ia from LSST are shown. Each row presents light curves over different redshift ranges, with the dotted line showing the depth of 4MOST after 1-hour integrated exposure (the maximum exposure for a single visit). The pink arrows denote the epoch/phase at which the object met the triggering requirements detailed in Section~\ref{sec:selectFunc}, and the black arrow indicates when 4MOST obtained the spectrum in our simulations. Note that even in our highest redshift bin, the uncertainties on the LSST observations are  small and constraining due to the superior depth of LSST \citep[$5\sigma\,$ detection, $r=24.3$][]{lsstSRD}. A SALT2 light-curve model (Section~\ref{sec:snIaLCFit}) was also fit to the observed light curves and we show the best fit result here.}
    \label{fig:exampleLightCurves}
\end{figure*}

One of the main objectives of TiDES is to perform a SN Ia cosmology experiment, so we next examine various survey performance metrics for SNe Ia. TiDES will obtain classification spectra for around \nIasnrFive\ SNe Ia with \snrF\,$\ge5$ (and \nIasnrThree\ SNe Ia \snrF\,$\ge3$). Randomly selected example light curves are shown in Fig.~\ref{fig:exampleLightCurves} for the $griz$ bands split into different redshift ranges of objects that were successfully observed by 4MOST.

\begin{figure}
    \centering
    \includegraphics[width=\columnwidth]{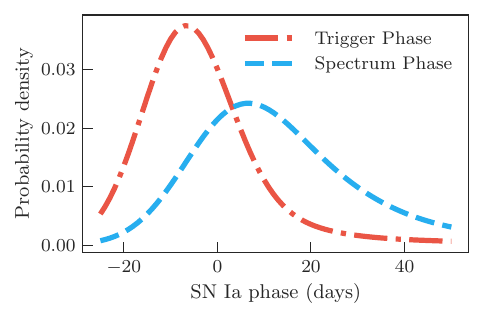}
    \caption{For the live-SNe Ia observed by 4MOST, we compare the phase at which we trigger follow-up requests to the phase the first spectrum was eventually taken in the rest-frame. In our simulations, the vast majority of our SNe Ia were identified and sent to 4MOST before maximum brightness. The spectra were taken most frequently around 1 week after peak.}
    \label{fig:phaseOfSpec}
\end{figure}

 The phase at which spectra are obtained is an important metric, as the ability to classify an object depends both on the phase and SNR of the spectrum. Objects observed near peak-brightness will have the highest SNR and classifiers will have large template libraries of historical observations to draw upon. In Fig.~\ref{fig:phaseOfSpec} we show the phases (measured relative to rest-frame $B$-band peak) when the SNe Ia are triggered for follow-up (Section~\ref{sec:selectFunc}) and of their spectral observations.  Most SNe are identified by LSST before they reach peak brightness, and then typically observed by 4MOST a few days after peak brightness. A break down of the triggering and spectrum phases for all transient types are presented in Figure~\ref{fig:phasesMultiple}.

\begin{figure}
    \centering
    \includegraphics[width=\columnwidth]{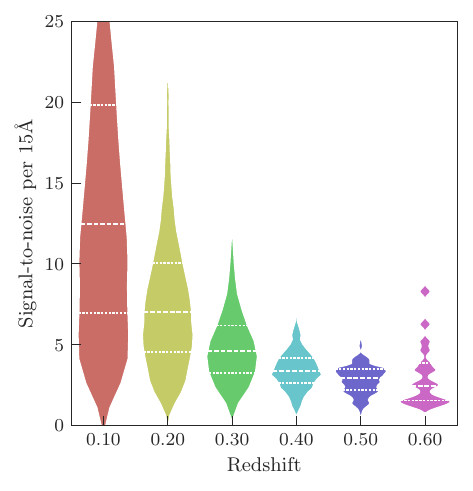}
    \caption{The distribution of the \snrF\, parameter for SNe Ia is shown and split across several different redshifts bins. The shape of the distribution is dictated both by the supernova phase at which the spectrum was acquired and by the exposure time of the 4MOST observation. In the lowest redshift bin, supernovae can be observed at their peak phases in longer exposures resulting in a higher SNR spectra, they can also be observed at later and fainter phases to produce the low SNR spectra. For the highest redshift bins, the SNe Ia are only observable at the peak brightness phase in the longest 4MOST exposures. Given the known uniformity of intrinsic brightness of SNe Ia at peak, this produces a narrower distribution, albeit at the lowest SNR. The dotted and dashed lines in each distribution represent the quartiles of the distribution.
    }
    \label{fig:SNRFuncOfRedshift}
\end{figure}

In the violin plot of Fig.~\ref{fig:SNRFuncOfRedshift}, we show the \snrF\, metric in different redshift bins.
We see broad distributions of \snrF\ in the lowest redshift bins that narrow with increasing redshift and fall to lower mean \snrF\ values. This is explained by the light curve evolution of our transients: at low redshifts the window-of-opportunity to observe a transient is large and  a spectrum can be obtained over many phases and SN brightnesses and hence \snrF. At higher-$z$, objects are only observed around the peak of their light curves with a narrower range in brightnesses, and hence the \snrF\ distribution narrows with a smaller mean value.

\begin{figure}
    \centering
    \includegraphics[width=\columnwidth]{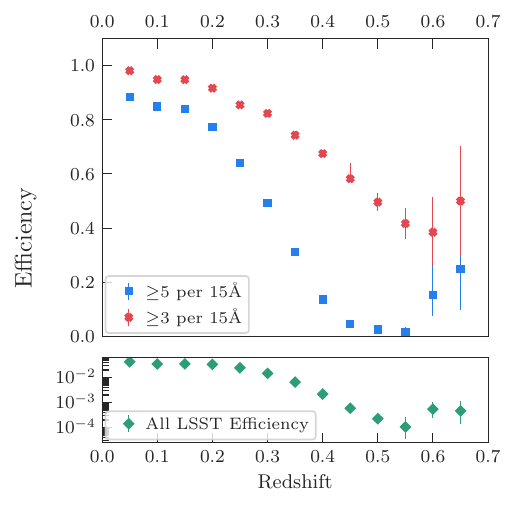}
    \caption{The fraction of spectra, split by \snrF\,$\ge3$ (red) and $\ge5$ (blue),  as a function of the object's redshift is shown for our SNe Ia sample. In the top panel, we show the efficiency for all SNe Ia 4MOST attempted to observe. In the bottom panel we show the fraction of all LSST SNe Ia that 4MOST successfully observed. While 4MOST-TiDES will provide a revolutionary sample of SNe Ia for cosmology, at best we will only be able to observe ${\sim3}\%$ of the SNe Ia that LSST discovers in any given redshift bin. Statistical-only uncertainties are shown following the same method as Figure~\ref{fig:effFuncMag}}
    \label{fig:fibreEff}
\end{figure}

Finally, we quantify the efficiency of obtaining a good-quality spectrum as a function of redshift. This metric analyses the success rate of the spectra meeting an SSC criteria and is visualised in Fig.~\ref{fig:fibreEff} (top panel). Objects at the lowest redshifts are almost always observed with \snrF\,$\ge3$, as these objects will be the brightest in the sample. Naturally, our \snrF\,$\ge5$\, curve falls away more rapidly, as spectra of this quality become increasingly more difficult to obtain in a fixed exposure time. 

The lower panel of Fig.~\ref{fig:fibreEff} shows the fraction of good-quality spectra for \textit{all} SNe Ia that we trigger spectroscopic follow-up requests for following the selection function in Section~\ref{sec:selectFunc}. As expected, the fraction of all LSST transients that TiDES will observe is small, highlighting the number of transients that Rubin will discover.

\subsection{TiDES host galaxies}
\label{sec:galaxy-results}

TiDES-Hosts is focused on spectroscopy of SN host galaxies. At the most fundamental level we aim to obtain secure redshifts for these hosts but, as exposure depth is built up for some targets, the increasing signal-to-noise will enable us to measure additional properties (e.g., star formation rate, gas phase metallicity \citep[][]{2023RASTI...2..453D}).

In TiDES, the galaxy spectra can be obtained on multiple epochs and then stacked to increase the signal-to-noise until the SSC in Section~\ref{sec:ssc} is reached. During 4MOST operations, automatic and manual redshifting will be performed on the new or stacked data products. Galaxies with prominent emission lines in their spectra can have their redshifts accurately determined before the continuum-defined SNR SSC are satisfied. A feedback-loop in the 4MOST system will remove such targets before additional fibre-hours are unnecessarily spent. However, this feedback is not incorporated in our simulations and instead observations are continued until the SSC is satisfied.

\begin{figure}
    \centering
    \includegraphics[width=\columnwidth]{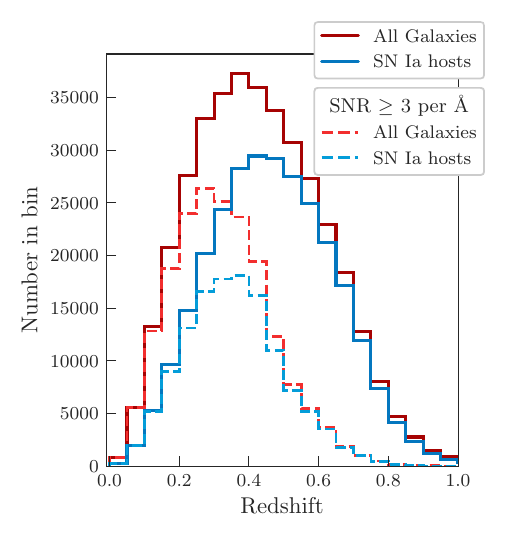}
    \caption{The redshift distribution of the host galaxies targeted by 4MOST-TiDES. The solid lines show the sample receiving at least some fibre-hours of observation, while the dashed lines show the sub-sample reaching \snrO\,$\ge 3$. Redshifts shown are the simulated galaxy redshifts.}
    \label{fig:nzGalaxy}
\end{figure}

We generate galaxy mock spectra as described in Section~\ref{sec:mockSpectra} and apply the technique to every galaxy (Section~\ref{sec:sim-hosts}) observed in the 4MOST-TiDES simulation. In total ${\sim}$\nGotHosts\, galaxies were targeted by 4MOST but not all satisfied our requirement of \snrO\,$\ge 3$ (Section~\ref{sec:ssc}) in the continuum. This is because the galaxies are injected during survey operations, resulting in fewer opportunities to add observations into a stacked spectrum as time progresses. In Figure~\ref{fig:nzGalaxy}, we show the redshift distribution split on both the total targeted by 4MOST those achieving \snrO\,$\ge 3$. Of these, our simulations show that ${\sim}$\nHostsIASNRthree\, hosted SNe Ia. We assume in later sections of our analysis, that all objects reaching our SSC will have a reliable redshift measured from the host galaxy spectrum. Next, we demonstrate a cosmology analysis from our combined results of live SNe Ia (Section~\ref{sec:live-sn-pop}) and from their host galaxy spectra.

\section{Supernova Cosmology}
\label{sec:SNIa-cosmology}

In this Section, we will present our expectations for the constraints TiDES can place on cosmological parameters following the methods employed by current generation experiments such as the Dark Energy Survey  \citep[][]{2024ApJ...973L..14D}. For this analysis, the spectroscopic classification and redshift measurement is assumed to be perfect from the TiDES-Live program, and objects without live-spectra but with host spectroscopic redshifts are also assumed to have a perfect photometric typing based on the light curves. 

The two SN Ia samples are based on the results of our LSST+TiDES simulations and are determined as follows:
\begin{itemize}
    \item \lq TiDES spectroscopic SN sample\rq: the sample of spectroscopically confirmed SNe Ia, for which TiDES has obtained one or more live spectra with a signal-to-noise ratio of ${\ge}3$ per 15\AA\ bin described in Section~\ref{sec:ssc}. Light curves for each SN are taken from the LSST photometry (Section~\ref{sec:sim_transients}).
    \item \lq LSST+TiDES photometric SN sample\rq: the sample of photometrically identified SNe Ia from their LSST light curves, for which the spectroscopic redshift of the SN host galaxy has been measured by TiDES following the \snrO\,$\ge 3$ in the continuum cut.
\end{itemize}

Following the forecasts presented in the Dark Energy Science Collaboration (DESC) Science Requirement Document \citep{2018arXiv180901669T}, we also include an external low-$z$ sample in our cosmology analysis of 2\,400 SNe Ia at $z{<}0.1$. The low-$z$ SNe in our analysis follows the simulations of the Foundation SN sample \citep{2018MNRAS.475..193F} presented by \citet{2019ApJ...881...19J}. The expectation, however, is that this final sample size will only be achieved towards the end of 10-years of LSST operations, even though TiDES is expected to conclude after the first five years of LSST.

\subsection{Defining the SN sample}
\label{sec:snIaLCFit}

We fit all simulated SNe Ia with the SALT2 light curve fitter \citep{2007A&A...466...11G} based on the simulated observed photometry. To ensure meaningful light curve fits, we follow standard requirements \citep[e.g.][]{2019ApJ...874..150B} of: i) observations in at least two different filters with SNR$>$5, ii) at least one data point before the time of peak brightness, $t_{\mathrm{peak}}$, and iii) at least one data point ten days after $t_{\mathrm{peak}}$. The first of these are already satisfied by the the spectroscopic triggering criteria in Section~\ref{sec:selectFunc}, and the final two are satisfied in the majority of cases due to the high cadence of WFD in LSST. 

Once the model has been fit to light curves, a final subset of the SNe Ia is selected by applying the following SALT2-based cuts \citep[e.g.,][]{2014A&A...568A..22B,2019ApJ...874..150B}:
$x_1 \in [-3,3]$, $c \in [-0.3,0.3]$, $\sigma_{x_1}<1$, $\sigma_{\text{peak}}<2$ days, and fit probability $>$0.001. Those that pass this step are considered ‘cosmologically useful’ supernovae. These selection criteria are generally adopted in SN Ia cosmology analyses to reduce contamination from highly-reddened or other peculiar thermonuclear SNe while ensuring a minimal loss of \lq cosmologically useful\rq\ SN Ia events. We find ${\sim}$\nCosmoLiveIa\, successful SALT2 fits in the spectroscopic sample and ${\sim}$\nCosmoHosts\, successful fits in the photometric SN sample. The redshift distribution for both samples and the low-$z$ sample are shown in the top panel of Fig.~\ref{fig:hubbleD}.

\subsection{SN distance estimation and bias corrections}

\begin{figure*}
    \centering
    \includegraphics[width=\linewidth]{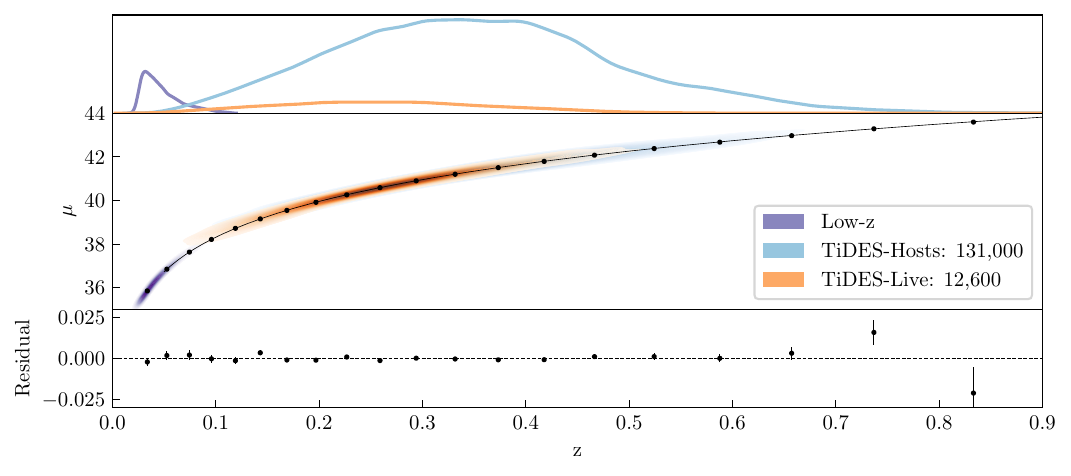}
    \caption{The TiDES Hubble Diagram made up of the final SN Ia sample presented in Section~\ref{sec:SNIa-cosmology}. The upper panel shows the redshift distribution of the TiDES-Live, TiDES-Hosts, and low-$z$ sample \citep{2019ApJ...881...19J}. The lower panel shows the residuals in the redshift bins from the cosmological fit (Hubble residuals).}
    \label{fig:hubbleD}
\end{figure*}

After light-curve fitting, we measure SN Ia distances and build the redshift--distance relation (\lq Hubble diagram\rq) from which cosmological parameters can be constrained. We estimate the SN distance modulus $\mu_\mathrm{obs}$, which is proportional to the logarithm of the SN luminosity distance, by applying the Tripp formula \citep[][]{1998A&A...331..815T}
\begin{equation}
    \mu_\mathrm{obs} = m_B + \alpha x_1 - \beta c - \mathcal{M}_B + \mu_\mathrm{bias},
    \label{tripp}
\end{equation}
where symbols are as defined in Section~\ref{sec:simSNeIa}. The nuisance parameters $\alpha$, $\beta$ and $\mathcal{M}_B$ are determined following the approach presented in \citet[][]{2011ApJ...740...72M}. We present our final Hubble diagram in Fig.~\ref{fig:hubbleD}.

The term $\mu_\mathrm{bias}$ in Equation~\ref{tripp} is a bias correction term that is applied to each SN to correct for survey selection effects. In this work, bias corrections are measured using the BEAMS with Bias Corrections \citep[BBC;][]{2019MNRAS.485.1171K} code. BBC uses large simulations of the survey to model and correct for selection effects while simultaneously fitting for the nuisance parameters and intrinsic SN scatter. For our analysis, corrections for selection effects are modelled as a function of redshift only (so-called \lq 1D\rq\ corrections) and they are estimated using the same LSST+TiDES simulated samples used to measure the cosmological constraints.

The output of BBC is a redshift-binned Hubble diagram corrected for selection effects, and the associated diagonal covariance matrix, $C_{\mathrm{stat}}$, that includes statistical uncertainties only for each redshift bin.

\subsection{Cosmological Parameters}

We use the output of the BBC fit and statistical+systematic covariance matrix to estimate cosmological contours for a Flat$w$CDM model (i.e., assuming a constant value of $w$), and for a Flat$w_0w_a$CDM model, where $w(a)=w_0+(1-a)w_a$ and $a$ is the scale factor.  
We estimate cosmological contours using \texttt{wfit.exe} $\chi^2$ minimization code implemented in \snana. 

\begin{figure}
    \centering
    \includegraphics[width=0.9\linewidth]{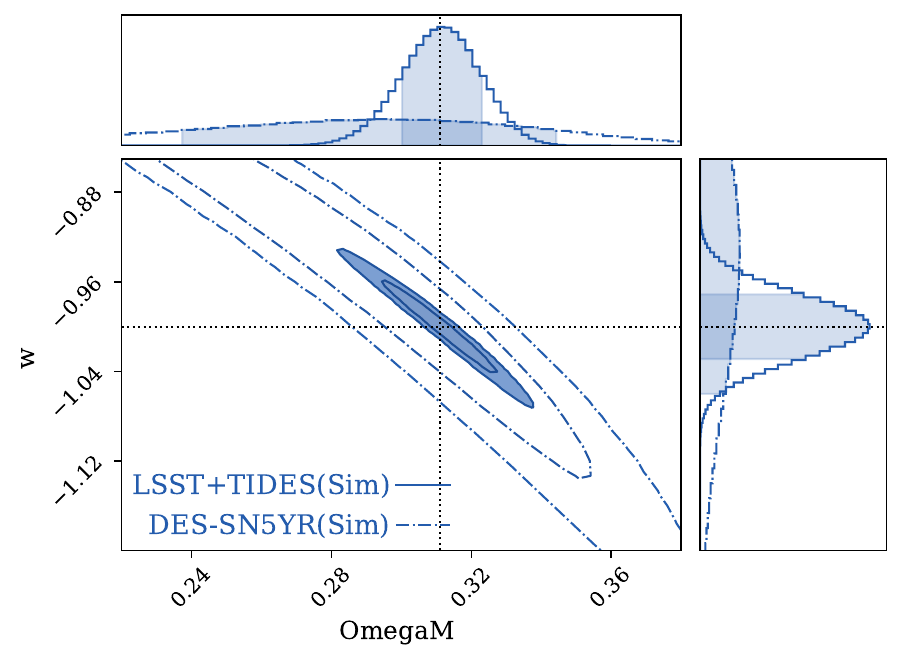}
    \caption{$\Omega_M - w$ contours for the simulated LSST+TiDES-Live and LSST+TiDES-Host combined (filled contours). For comparison, we present the cosmological contours from a DES-SN5YR\textit{-like} simulation (dotted-dashed contours). Only statistical uncertainties are included in both the LSST+TiDES and DES\textit{-like} contours.}
    \label{fig:cosmo_Omegamw}
\end{figure}

\begin{figure}
     \centering
     \includegraphics[width=0.9\linewidth]{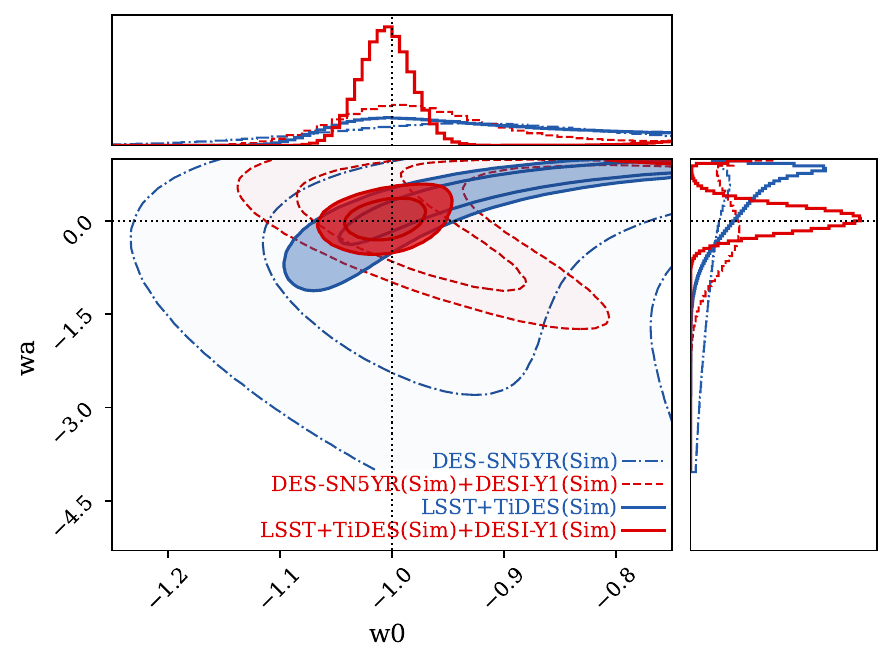}
     \caption{As Fig.~\ref{fig:cosmo_Omegamw}, but showing constraints in the $w_a$/$w_0$ plane. Additionally, we show the contours combining SN data with a DESI-BAO-Y1\textit{-like} prior, both for DES-SN5YR-like simulation (red dashed line) and LSST+TiDES simulation (red solid line).}
     \label{fig:cosmo_w0wa}
 \end{figure}

The cosmological contours from our simulations are presented in Fig.~\ref{fig:cosmo_Omegamw} and Fig.~\ref{fig:cosmo_w0wa}. We present constraints from the LSST+TiDES SN events (both spectroscopically-confirmed SNe Ia and SNe Ia with host spec-$z$ only) and the external low-$z$ sample. 
We compare our results with constraints from a simulation of the Dark Energy Survey 5 year SN sample \citep[DES-SN5YR, ][]{2024ApJ...973L..14A}. The DES-SN5YR analysis constitutes the state-of-the-art of SN cosmological measurements and provides some of the tightest constraints on the Dark Energy Equation of State. The DES-SN5YR simulation used in this work is generating assuming the same cosmology used in our LSST simulations (i.e., Flat$\Lambda$CDM cosmology with $\Omega_M=0.315$) and reflects the statistical power and data quality of the real DES-SN5YR data. While the published DES-SN5YR include statistical and systematic uncertainties, in Fig.~\ref{fig:cosmo_Omegamw} and Fig.~\ref{fig:cosmo_w0wa} we only consider statistical uncertainties. For a Flat$w$CDM model, LSST+TiDES SN samples combined with external low-$z$ samples and a CMB prior \citep[][]{2020A&A...641A...6P} provide constrains on the Dark Energy Equation of State parameter, $w$, of 0.012, 10 times smaller compared to DES-SN5YR. For a Flat$w_0w_a$CDM model, the Figure of Merit obtained from LSST+TiDES SN samples combined with external low-$z$ samples is 85, 15 times larger than the FoM of DES-SN5YR. When fitting for a Flat$w_0w_a$CDM model, we also combine SN data with a prior based on the recent Baryonic Acoustic Oscillation (BAO) measurements published by the Dark Energy Spectroscopic Instrument (DESI) Collaboration \citep{2025JCAP...02..021A} using their first year of data. The DESI-BAO-Y1 prior is applied assuming the same true cosmology used to generate the LSST and DES-SN5YR simulations. With the DESI-BAO-Y1 prior, the LSST+TiDES FoM increases to 99 (7 times larger compared to DES-SN5YR combined with the same DESI-BAO-Y1 prior).

\section{Summary}
\label{sec:summary}

We have presented the Time-Domain Extragalactic Survey (TiDES) -- a  spectroscopic survey focused on understanding the extragalactic transient universe, and conducted on the 4m Multi-Object Spectroscopic Telescope (4MOST). TiDES will use 250\,000 fibre-hours to address three key science goals: i) Create a Hubble Diagram of ${>}$\nAllCosmoAbout\, cosmologically-useful SNe Ia to constrain cosmological measurements, including the dark-energy equation-of-state, $w$, to sub-2 per cent precision; ii) Map the diversity of the transient Universe by obtaining more than \nSNeAbout\, spectra of supernovae from LSST; iii) Perform a reverberation mapping experiment on 700--1\,000 AGN in the high-cadence Deep-Drilling Fields.

In this work, we have simulated both the Rubin LSST and 4MOST surveys using the latest assumptions on their respective strategies. This has allowed us to create realistic light curves for millions of LSST transients with mock spectra for the sub-set that 4MOST can observe. Our results demonstrate that we will collate the largest homogeneous sample of supernovae and their host-galaxies to-date. Our results are summarised as follows:

\begin{itemize}
    \item We will obtain ${\sim}$\nSNeAbout\, SN spectra with a signal-to-noise ratio ${\ge}3\,$ per 15\AA\, (\snrF\,).  Of these, ${\sim}$\nCosmoLiveIa\, will be type Ia supernovae with a light curve quality suitable for a cosmological analysis.
    \item TiDES will observe ${>}$\nCCThree\, core-collapse supernovae and create a sample of rare faint-and-fast transients larger than currently exists from any other survey to-date.
    \item The TiDES-Hosts follow-up program will observe a sample of ${>}\nHostsSNRthree\,$ galaxies, of which ${\sim}\nCosmoHosts\,$ will host SNe Ia suitable for cosmology.
    \item Our cosmologically-useful sample, combining live-SNe Ia and photometrically identified SNe Ia with host galaxy redshifts, will be placed on a Hubble Diagram containing at-least \nAllCosmo\, objects.
    \item We performed a cosmology analysis of the LSST+TiDES sample, including a low-$z$ sample predicted by \citet{2019ApJ...881...19J}. We are able to constrain the dark energy equations-of-state parameter, $w$, to a sub-2 per cent statistical-only precision, this is 10 times smaller compared to the current gold-standard \citep{2024ApJ...973L..14D}.
    \item When we consider our constraints on a Flat$w_0w_a$CDM model, our LSST+TiDES SN sample produces a figure-of-merit value 15 times larger than DES-SN5YR.
    
\end{itemize}

The Rubin Observatory's Legacy Survey of Space and Time will transform our understanding of the time-domain Universe. Important in unlocking the potential of these data will be the spectroscopic follow-up of transients, their host galaxies, and of active galactic nuclei. TiDES is primed to address this challenge providing dedicated follow-up time across the 5-year survey. All data collected by TiDES and other participating 4MOST surveys will be made public in periodic data releases on the ESO Science Archive Facility.

\software{ NumPy \citep[][]{harris2020array}, Pandas \citep[][]{reback2020pandas,mckinney-proc-scipy-2010}, DuckDB \citep[][]{duckdb}, AstroPy \citep[][]{astropy:2013,astropy:2018,astropy:2022}, Matplotlib \citep[][]{Hunter:2007} }
\begin{acknowledgments}
The authors would like to thank Ricardo Demarco and Swayamtrupta Panda for their comments on this manuscript.

This paper has undergone internal review in the LSST Dark Energy Science Collaboration. The internal reviewers were Chris Lidman and Alex Gagliano.

CF and MSu acknowledge support from STFC funding for UK participation in LSST, through grants ST/V002031/1 and  ST/X00130X/1. 
MN is supported by the European Research Council (ERC) under the European Union’s Horizon 2020 research and innovation programme (grant agreement No.~948381) and by UK Space Agency Grant No.~ST/Y000692/1.
SFH acknowledges funding from STFC Grant ST/Y001656/1 and from UK Research and Innovation (UKRI) under the UK government’s Horizon Europe funding Guarantee (EP/Z533920/1).
SJS acknowledges funding from STFC Grants ST/Y001605/1, ST/X001253/1, ST/X006506/1, ST/T000198/1 and the a Royal Society Research Professorship grant.
AM is supported by the ARC Discovery Early Career Researcher Award (DECRA) project number DE230100055.
Y.-L.K. has received funding from the Science and Technology Facilities Council [grant number ST/V000713/1]. Y.-L.K. was supported by the Lee Wonchul Fellowship, funded through the BK21 Fostering Outstanding Universities for Research (FOUR) Program (grant No. 4120200513819) and the National Research Foundation of Korea to the Center for Galaxy Evolution Research (RS-2022-NR070872, RS-2022-NR070525).
LK acknowledges support for an Early Career Fellowship from the Leverhulme Trust through grant ECF-2024-054 and the Isaac Newton Trust through grant 24.08(w).
ET acknowledges funding from the HTM (grant TK202), ETAg (grant PRG1006) and the EU Horizon Europe (EXCOSM, grant No. 101159513).\\

The DESC acknowledges ongoing support from the Institut National de 
Physique Nucl\'eaire et de Physique des Particules in France; the 
Science \& Technology Facilities Council in the United Kingdom; and the
Department of Energy, the National Science Foundation, and the LSST 
Corporation in the United States.  DESC uses resources of the IN2P3 
Computing Center (CC-IN2P3--Lyon/Villeurbanne - France) funded by the 
Centre National de la Recherche Scientifique; the National Energy 
Research Scientific Computing Center, a DOE Office of Science User 
Facility supported by the Office of Science of the U.S.\ Department of
Energy under Contract No.\ DE-AC02-05CH11231; STFC DiRAC HPC Facilities, 
funded by UK BEIS National E-infrastructure capital grants; and the UK 
particle physics grid, supported by the GridPP Collaboration.  This 
work was performed in part under DOE Contract DE-AC02-76SF00515.

\end{acknowledgments}
\begin{contribution}
CF led the analysis, performed the simulations, and wrote the paper. MV and MSm performed the cosmology analysis. MSu discussed the results, implications, and commented on the manuscript at all stages of the project. SFH is PI of the TiDES-RM program and wrote the relevant sections of this paper. All other co-authors are members of 4MOST/TiDES and have commented on this work.\\
\end{contribution}

\appendix

\section{Trigger and Spectrum Phases}

\begin{figure}
    \centering
    \includegraphics[width=0.5\linewidth]{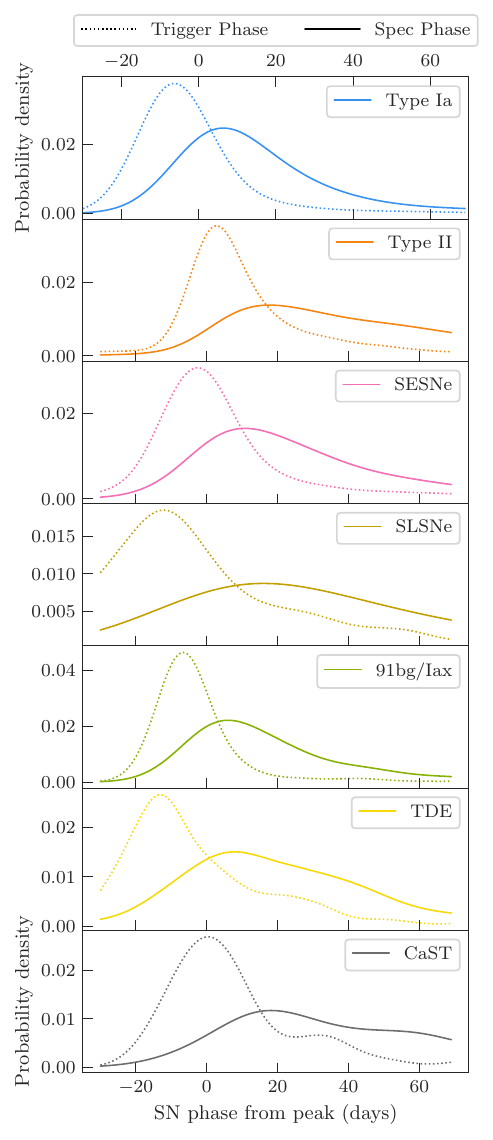}
    \caption{The distribution of light curve trigger phases and the phase of spectroscopic follow up by 4MOST is shown, broken down into transient sub-types. All phases are measured relative to the light curve peak. The methodology used to generate this figure is identical to that presented in Figure~\ref{fig:phaseOfSpec}.}
    \label{fig:phasesMultiple}
\end{figure}

\bibliography{references}{}
\bibliographystyle{aasjournal}



\end{document}